\documentstyle[eepic,12pt]{article}
\renewcommand{\baselinestretch}{1.3}

\begin{document}


\newfont{\abg}{cmsl12}
\def\gtrue{\mbox{{\abg {\bf g}}}} \def\otrue{\mbox{{\abg {\bf $\o$}}}}
\newcommand{\pref}[1]{\ref{#1}\fbox{#1}}                      
\newcommand{\plabel}[1]{\label{#1}\fbox{#1}}                  
\newcommand{\prefeq}[1]{Gl.~(\ref{#1}\fbox{#1})}              
\newcommand{\prefb}[1]{(\ref{#1}\fbox{#1})}                   
\newcommand{\prefapp}[1]{Appendix~\ref{#1}\fbox{#1}}          
\newcommand{\plititem}[1]{\begin{zitat}{#1} {\fbox{#1}}a
                                             \end{zitat}}     
\newcommand{\plookup}[1]{\hoch{\ref{#1}}\fbox{#1} }           
\newcommand{\pcite}[1]{\cite{#1}\fbox{#1} }                  
\newcommand{\hinweis}[1]{\fbox{#1}}                           
\newcommand{\BM}{B$\,\&$M$\:\,$}


\def\Di{\displaystyle}
\def\nn{\nonumber}
\def\sr{\stackrel}
\def\be{\begin{equation}}
\def\ee{\end{equation}}
\def\ba{\begin{eqnarray}}
\def\ea{\end{eqnarray}}
\def\pl{\plabel} \def\lab{\label}
\def\re{(\ref }
\def\rz#1 {(\ref{#1}) }   \def\ry#1 {(\ref{#1})}
\def\el#1 {\plabel{#1}\end{equation}}

\let\a=\alpha \let\b=\beta \let\g=\gamma \let\d=\delta
\let\e=\varepsilon \let\ep=\epsilon \let\z=\zeta \let\h=\eta
\let\th=\theta
\let\dh=\vartheta \let\k=\kappa \let\l=\lambda \let\m=\mu
\let\n=\nu \let\x=\xi \let\p=\pi \let\r=\rho \let\s=\sigma
\let\t=\tau \let\o=\omega \let\c=\chi \let\ps=\psi
\let\ph=\varphi \let\Ph=\phi \let\PH=\Phi \let\Ps=\Psi
\let\O=\Omega \let\S=\Sigma \let\P=\Pi
\let\Th=\Theta \let\L=\Lambda \let \G=\Gamma \let\D=\Delta
\def\q{\quad} \def\qq{\qquad}
\def\wt{\widetilde}
\def\w{\wedge}
\def\0{\over} \def\1{\vec} \def\2{{1\over2}} \def\4{{1\over4}}
\def\5{\bar} \def\6{\partial}

\def\({\left(} \def\){\right)} \def\<{\langle} \def\>{\rangle}
\def\lb{\left\{} \def\rb{\right\}}
\def\[{\left[} \def\]{\right]}
\let\lra=\leftrightarrow \let\LRA=\Leftrightarrow
\def\ul{\underline}
\let\Ra=\Rightarrow \let\ra=\rightarrow
\let\la=\leftarrow \let\La=\Leftarrow

\let\ap=\approx \let\eq=\equiv \let\hc=\dagger
\let\ti=\tilde \let\bl=\biggl \let\br=\biggr
\def\CL{\cal L} \def\CX{\cal X} \def\CA{\cal A}
\def\CF{\cal F} \def\CD{\cal D} \def\rd{\rm d}
\def\rD{\rm D} \def\CH{\cal H} \def\CT{\cal T} \def\CM{\cal M}
\def\CI{\cal I} \newcommand{\dR}{\mbox{{\rm I \hspace{-0.8em} R}}}
\newcommand{\dN}{\mbox{{\sl I \hspace{-0.8em} N}}}
\def\P{\cal P} \def\CS{\cal S} \def\C{\cal C}
\def\bC{{\bar{\C}}}
\def\CR{{\cal R}}\def\CO{{\cal O}}

\begin{titlepage}
\renewcommand{\thefootnote}{\fnsymbol{footnote}}
\renewcommand{\baselinestretch}{1.3}
\hfill  PITHA -- 98/27\\
\medskip
\vspace{1mm}
\begin{center}
{\Large{BRST Inner Product Spaces and the Gribov Obstruction}} 

\vspace{1cm}
Norbert D\"UCHTING$^1$\footnote{email:
norbertd@physik.rwth-aachen.de}, Sergei V.\ 
SHABANOV$^2$\footnote{email: shabanov@phys.ufl.edu}\footnote{On leave 
from the Laboratory of Theoretical Physics, JINR, Dubna, Russia}, Thomas
STROBL$^1$\footnote{email: tstrobl@physik.rwth-aachen.de}\\

\vspace{0.7cm}
$^{1}\hspace{3mm}$ Institut f\"ur Theoretische Physik, RWTH Aachen\\
D-52056 Aachen, Germany\\
$^2\hspace{3mm}$ Department of Mathematics, University of Florida,\\ 
Gainesville, FL 32611--8105, USA 

\medskip

\begin{abstract}
A global extension of the Batalin--Marnelius proposal for a BRST
  inner product to gauge theories with topologically nontrivial gauge
  orbits is discussed. It is shown that their (appropriately adapted)
  method is applicable to a large class of mechanical models with a
  semisimple gauge group in the adjoint and fundamental
  representation.  This includes cases where the Faddeev--Popov method
  fails. Simple models are found also, however, which do {\em not}\/
  allow for a well--defined global extension of the Batalin--Marnelius
  inner product due to a Gribov obstruction. Reasons for the partial
  success and failure are worked out and possible ways to circumvent the
  problem are briefly discussed.
\end{abstract}
\end{center}
\end{titlepage}

\section{Introduction and Overview}

Canonical quantization of gauge theories leads, in general, to an
ill--defined scalar product for physical states.  In the Dirac
approach \cite{dirac} physical states are selected by the quantum
constraints. Assuming that the theory under consideration is of the
Yang--Mills type with its gauge group acting on some configuration space,
the total Hilbert space may be realized by square integrable functions
on the configuration space and the quantum constraints imply gauge
invariance of these wave functions. Thus the physical wave functions
must be constant along the orbits traversed by gauge transformations
in the configuration space. Consequently, the norm of the physical
states is infinite, if the gauge orbits are noncompact or if the number
of nonphysical degrees of freedom is infinite as in gauge {\em field}\/
theories.

The problem is similar to that which occurs also in the path integral
quantization of gauge theories where the integral over gauge field
configurations diverges because of the gauge invariance of the action.
In their seminal work \cite{FP} Faddeev and Popov proposed a solution
based on the idea that in order for the path integral measure to be
finite, only one representative of each gauge orbit should be taken
into account. They provided a systematic way of implementing gauge
conditions in the integral so that, by inclusion of an appropriate
additional contribution to the measure, namely the Faddeev--Popov (FP)
determinant, the resulting integral becomes independent of the choice
of gauge and effectively ranges over the physical degrees of freedom
only. However, if the gauge orbits possess a nontrivial topology, as
often happens in physically interesting theories, a good choice of
gauge turns out to be impossible. This deficiency is known as the
Gribov obstruction \cite{Gribov,singer,jones}. It can be illustrated
with simple mechanical models \cite{ufn} that ignoring global
deficiencies of a particular gauge can result in explicitly wrong
predictions of the corresponding path integral quantization.

The FP path integral measure specifies uniquely the measure in
the scalar product. In fact, the norm of the Dirac states
can be made finite by reducing the initial measure to the
gauge fixing surface and inserting the corresponding FP determinant
to maintain the gauge invariance of the physical amplitudes.
The Gribov obstruction will again be present,  if the gauge
orbits have a nontrivial topology. So, the FP approach would,
in general, lead to an ill--defined scalar product.
In many of these cases, one may, however, further modify the resulting FP
inner product so as to obtain a finally reasonable norm for physical
states. In the simplest case this is effected, e.g., by restricting
the domain of integration along the gauge fixing surface to its
modular domain (i.e.\ to a region which contains no points that are
still gauge equivalent to others on that surface, their so--called
Gribov copies).

More recent suggestions to handle the norm regularization problem
within the Dirac approach include a redefinition of the scalar product
along the lines suggested in \cite{zedon,land} or the transition to
the coherent state representation for constrained systems as performed
in \cite{kl97,klsh97}.

In gauge field theories and especially in their path integral
formulation, an explicit Lorentz invariance of the quantum theory is
desired, which is not available in the Dirac Hamiltonian approach. The
Lorentz invariance can, however, be achieved within the (nonminimal)
BRST quantization program (see, e.g., \cite{marc}).

BRST quantization is based on the extension of the original phase
space by Lagrange multiplier and ghost sectors. For a set of
first--class constraints $G_a$ one introduces the Lagrange multipliers
$y^a$ and their canonical momenta $p_{y^a}$ and adds canonical pairs
$({\cal C}^a,{\cal P}_a)$ and $(\bar{\cal C}^a,\bar{\cal P}_a)$ of
fermionic (Grassmann) ghost and antighost variables, respectively.
The extended phase space has a natural grading with respect to the
ghost number operator $N$: $[N,{\cal C}^a]={\cal C}^a,\ [N,\bar{\cal
  C}^a]=-\bar{\cal C}^a$, etc.  Finally, the BRST charge $Q$ is
constructed. It is a hermitian nilpotent operator ($Q^\dagger =Q,
Q^2=0$) such that, at least generically, the Dirac physical subspace
formed by functions on the orbit space is isomorphic to the subspace
composed of elements of BRST--cohomology classes (usually at ghost
number zero).  That is, one looks for wave functions that are
annihilated by $Q$, called BRST--closed, and identifies those
differing by elements of the image of this operator (identification
by BRST--exact states).  Formally different representatives chosen
from an equivalence class yield the same physical answers since \be \<
s_1 | \( |s_2\> + Q |p\> \) = \< s_1|s_2\> \, \, , \ee where we made
use of  $Q|s_1\>=0$ and  the hermiticity of $Q$.  However, it turns
out that in practice the physical states (among others) often do not
have a well--defined norm in the original (indefinite) Hilbert space.
Typically, the norm is proportional to the meaningless factor
$\infty\cdot 0$. The infinity comes from the integration over the
Lagrange multipliers, while zero results from the Berezin integral
over the ghosts and antighosts (cf., e.g., \cite{marc} or Sec.\ 2
below, providing a simple illustration). So, such as in Dirac
quantization, also in canonical BRST quantization there is a problem
with the inner product \cite{cobrst}, which, moreover, has not been
resolved  in generality up to present day.  

In this work we shall discuss an approach due to Batalin and Marnelius
(\BM) to this problem \cite{M1}. Their main idea is to {\em not}\/
alter the original inner product, but to single out specifically
chosen representatives in the BRST cohomology classes which then have
a well--defined inner product among each other. They provide a scheme
to construct a (hermitian) gauge fixing fermion $\Psi$ and a space of
so--called auxiliary states $|s\>_0$, which, as we will see, resemble
the physical (gauge invariant) states obtained in the ghost--free
Dirac approach. The representatives of the cohomology which have a
well--defined inner product are then provided by the BRST singlets
\be
|s\> := e^{[Q,\Psi]_+}|s\>_0\ .
\label{1a}
\ee 
More precisely, Batalin and Marnelius
{\em define}\/ the inner product of physical states, which are in a
one--to--one correspondence with the states $|s\>_0$, by means of
\be
\<s|s'\> = {}_0\<s| \exp \( 2[Q,\Psi]_+\) | s'\>_0 \, ,
\label{master1}
\ee which follows {\em formally}\/ from the above representation of
$|s\>$ and the analogous one for $|s'\>$ when using (naive)
hermiticity of $Q$ and $\Psi$. Note that the auxiliary states $|s\>_0$
have a specific dependence on the ghost and nonphysical variables. So
they are also called the ghost-- and gauge--fixed states. The BRST
transformed states (\ref{1a}) with a {\em generic} $\Psi$ yield, at
least up to global issues, the whole cohomology {\em class}\/
represented by $|s\>_0$.  The conventional BRST inner product between
quantum states is not always well--defined.  The goal of Batalin and
Marnelius was to develop a formalism for selecting a set of
representatives $|s\>_0$ together with an adapted $\Psi$ such that the
resulting representatives $|s\>$ have a well--defined inner product
with one another. The reason for introducing the states $|s\>_0$ on an
intermediate level is that generically they are much simpler than the
states $|s\>$, containing, e.g., no ghosts in an appropriate
polarization and sometimes even coinciding literally with the states
found in a Dirac quantization (cf.\ also the examples below in this
paper).

Arriving at formula \re{master1})
in this way, one implicitly has defined an inner product between the
BRST cohomology {\em classes}\/ (represented by $|s\>_0$ or $|s\>$).
The \BM solution (\ref{1a}) of the BRST inner product
cohomologies is local \cite{M1}. So the question of a global
extendibility of their formalism arises.

As follows from (\ref{master1}), the choice of gauge conditions ---
or, equivalently, the choice of the gauge fixing fermion $\Psi$ --- is
an essential ingredient of the \BM approach, 
such as it is in the FP procedure for
defining an inner product between Dirac quantum states. 
In gauge theories with a nontrivial topology of the gauge
orbits in the configuration space, there is, in general, a Gribov
obstruction to a (globally well--defined) choice of a gauge condition,
which can cause serious deficiencies of the FP inner product.
Therefore one might expect an analogous problem within the \BM
procedure.

The aim of the paper is to investigate possible global obstructions to
the \BM construction that might occur through a non--Euclidean gauge
orbit space geometry \cite{babelon}, which has not yet been addressed.
In the \BM for\-mal\-ism, one of the conditions placed on $\Psi$ and
$|s\>_0$ is that the FP determinant of the gauge underlying the choice
of $\Psi$ is nonvanishing everywhere. In the presence of the Gribov
obstruction this condition cannot be met anymore. We will study
possible consequences of fulfilling this condition only almost
everywhere in the configuration space, i.e.\ the associated FP
determinant is nonzero everywhere in configuration space except for
some region of lower dimension. It then will turn out that in some
cases the \BM method still provides a useful recipe for constructing
an inner product between BRST cohomologies, while in others it will
not.  Among these cases we will find examples where the FP method
fails, while the \BM procedure works.

To single out the crucial difference between gauge systems with and
without the Gribov problem, we shall apply the \BM procedure to simple
mechanical gauge models in which the Gribov obstruction is evident and
compare them with similar models where the latter is absent.  The
models chosen are simple enough to be analyzed by less sophisticated
methods to full accuracy, thus allowing for a first test of the \BM
version of the BRST approach. In particular, the configuration space
of the models will be finite dimensional always in this paper (which
is in contrast to field theories) and, moreover, in most (but not all)
of the models, the gauge group is compact and of finite volume.
Correspondingly, in all those miniature models with a compact gauge
group, irrespective of the 
presence of a Gribov problem, the physical wave functions in the Dirac
quantization will have a well--defined, nonsingular, and physically
reasonable inner product already with respect to the measure defined
in the {\em original}\/ Hilbert space. So, in {\em these}\/ cases the 
FP method (or similarly the \BM method) for constructing an inner
product is superfluous. However, we are still free to apply these
methods also to such simple models  and then compare the result to the
one obtained by the Dirac procedure, which we then may use as the
touchstone for a correct inner product, at least up to unitary
equivalence. 

In the following section (Sec.\ 2), we will briefly recapitulate the
idea of the \BM construction of the inner product at the example of
the simplest possible ``gauge theory'' without Gribov obstruction, the
gauge orbits being straight lines in a two--dimensional Euclidean
configuration space.  In Sec.\ 3 we apply the \BM recipe to a model
with gauge group $SO(2)$, the gauge orbits of the previous model
having been compactified to circles. The seemingly small deficiency of
a vanishing FP determinant at a set of zero measure in the
configuration space, which is a consequence of the Gribov problem,
turns out to be a decisive obstruction to the \BM method in this case
(for a specific choice of the gauge fixing fermion the inner product
vanishes identically).  Studying, on the other hand, the analogous
model with gauge group $SO(3)$ in Sec.\ 4, the gauge orbits being
spheres in an $\dR^3$ now, the \BM procedure is found to provide a
well--defined inner product, equivalent to the covariant result of
Dirac quantization.  To be precise, in order for the latter statement
to hold, some additional new condition in the construction of the BRST
operator $Q$ has to be met, which is not present in the work of
Batalin and Marnelius. Yet, also the gauge fixing fermion is to be
restricted in a certain way, discussed in more detail further below.

In Sec.\ 5 we then see that the successful application of the \BM
ap\-proach to the $SO(3)$--model can be extended to models with an
arbitrary semisimple Lie group acting in its adjoint representation.
The condition on the operator $Q$ will be specified and further
clarified in this context.

Much of the remainder of the paper is then devoted to the question,
why the proposal of Batalin and Marnelius, refined by 
the aforementioned condition on
the form of $Q$, works for the models
discussed in Sec.\ 4 and 5, while it fails for the simple $SO(2)$--
model.

The first and most near--at--hand ansatz to answering this question is
the following observation: For the models studied up to that point,
the FP method\footnote{For reducible theories (discussed further
below)  the FP determinant
is defined with respect to a subset of locally independent
constraints.}  works and fails in precisely the same cases as the \BM\ 
method does. This is not the full answer, however. As we will show in
Sec.\ 6.1, there are models for which the FP method fails, while the
\BM approach still works! These models are obtained from another
generalization of the $SO(3)$--model: Interpreting the action of the
$SO(3)$ group on the configuration space $\dR^3$ not as the adjoint
action in the Lie algebra as in the generalization of Sec.\ 5 (in which
case the configuration space 
variables are somewhat similar to gauge fields in realistic
Yang--Mills theories) but as the fundamental action, it is most
straightforward to generalize the $SO(2)$-- and $SO(3)$--model
simultaneously to obtain a model with gauge group $SO(N)$ acting in its
fundamental representation on $\dR^N$. This is interesting because it
may be seen that the FP method works for odd $N$ while it fails for
even $N$, producing a gauge dependent norm in the latter instance
which, in the worst case,  may even vanish. All the more it comes
somewhat as a surprise that the (appropriately refined) \BM method
yields a good inner product (equivalent to ``the correct'' one in the
original $N$--dimensional configuration space) provided only that $N
\ge 3$.

Another obvious difference between $SO(2)$ on the one hand and $SO(3)$
with all its successful generalizations on the other hand is that the
gauge group of the former model is abelian, while the gauge groups of
all the other models are semisimple, which, from a group theoretical
point of view (cf., e.g., \cite{Gilmore}), is something like the
extreme opposite of abelian. The model studied in Sec.\ 6.2 provides
an example to this guess demonstrating the opposite: Considering more
than just one particle in a three--dimensional configuration space
with the rotational group $SO(3)$ acting on all of them {\em
  simultaneously}\/, the \BM procedure is found to fail again. 

{}From all of these studies it appears to us that it could be the {\em
  reducibility}\/ of the constraints that allows for a successful
application of the \BM construct, while theories with irreducible
constraints generically will lead to unacceptable results in the
presence of a Gribov obstruction. Here reducibility of (first--class)
constraints $G_a \approx 0$ means that they are not independent from
one another, i.e.\ there exists at least one relation $Z^a G_a \equiv
0$ for some functions $Z^a$ on the phase space of the theory. Clearly
any theory formulated in an irreducible manner can be reformulated by
means of a reducible set of constraints. So, in the above,
``reducibility'' should be specified to what one might
call ``essential reducibility'', by which we mean a constrained
Hamiltonian system with reducible constraints which cannot be
replaced {\em globally}\/ by a set of irreducible first--class
constraints. The prototype of such a theory is the initially mentioned
$SO(3)$--model, the constraints being the three components of the
angular momentum in the phase space $T^*(\dR^3)$. The reducibility of
the $SO(3)$--model is
lost when the number of particles is increased above one, as is done
in Sec.\ 6.2.

We remark at this point that the condition mentioned above to refine
the \BM version of BRST quantization is one placed on the functions
$Z^a$ expressing the mutual dependence of the constraints. Classically
there is a large ambiguity or freedom in choosing such functions. Only
an appropriately chosen subset of those functions will lead to a good
inner product of the quantum theory, while others turn out to be
unacceptable in the end. 


For the models of Sec.\ 4 and 5, with the (semisimple and compact)
gauge group acting in the adjoint representation, we observe that the
(refined) \BM construction yields a BRST inner product that reproduces
the one found in the Dirac quantization (which, as remarked already
above, is also well--defined in these mechanical toy models). This
result holds at least for all choices of the BRST gauge fixing fermion
$\Psi$ which correspond to a gauge that is linear in the configuration
space variables. As the Dirac inner product certainly is independent
of any possible choice of gauge, we may conclude also
gauge independence of the \BM procedure, at least within the above
class of gauge fixing fermions. However, this apparent gauge
independence is by no means complete.  In Sec.\ 7 we will see at the
example of the helix model \cite{kl97,helix1,helix2,fujikawa} 
that the \BM procedure
fails for gauge conditions with a nonconstant number of Gribov copies.
Moreover, it is found that if the number of Gribov copies diverges ---
in Sec.\ 8 we will argue by means of an example why this is of
relevance in physical theories ---, then also the \BM inner product
diverges! This effect was not observed in Secs.\ 4 and 5 since there
the number $N_W$ of copies was finite; the resulting inner product
comes out proportional to $N_W$, but any finite proportionality
constant drops out from an inner product by normalization.

Sec.\ 8 contains our conclusions and a discussion of possible ways to
circumvent the global topological obstructions for constructing the
proper BRST inner product in physical theories.


\section{The Batalin--Marnelius procedure:  a simple example}


In this section we illustrate the \BM procedure \cite{M1}  by means of
a simple
example, namely a particle on ${\rm I\! R}^2$ parameterized by coordinates $x$
and $y$ with the translational gauge symmetry along the $y$-direction.
The gauge orbits are straight lines parallel to the $y$-axis.
So there is no Gribov problem, although the gauge
orbits have infinite volume, thus indicating already the 
divergence of a naive BRST scalar product. 
The model will allow us to recall the
main idea and ingredients of the \BM construction. 
Denoting the
momenta by $p_x$ and $p_y$, the constraint is simply $p_y \approx 0$,
while the Hamiltonian $H$ has to be independent of $y$ due to the
translational symmetry. 
The {\em nonminimal}\/ BRST charge is $Q=p_y \, {\cal C} +
\bar{\cal P} \, \pi$, where $(\cal C, \cal P)$ and $({\bar{\cal C}},
 \bar{\cal P})$ are
canonical pairs of fermionic ghost and antighost variables,
respectively, and $\pi$ is the momentum conjugate to the Lagrange
multiplier $\l$, which enforces the constraint within the Hamiltonian
action. The {\em nonminimality}\/ of the BRST scheme means that
the canonical pair $(\l,\pi)$, supplemented by the additional
first--class constraint $\pi \approx 0$, is added to the phase space.
The reason for doing this is the analogue with gauge field
theories where the addition of the Lagrange multipliers
to the BRST multiplet allows for explicit Lorentz covariance
(in contrast to the {\em minimal}\/ Hamiltonian approach where
the Lagrange multipliers are excluded before quantization).

The BRST invariant quantum states, $Q|\psi\>=0$, modulo
shifts on $Q$-exact states, $|\psi\> \rightarrow |\psi\>
+ Q|\phi\>$,  form a space that is isomorphic to the
Dirac physical subspace determined by the gauge invariance
condition $p_y|\psi\>=0$. Since zero eigenvalue of the operator
$p_y$ lies in the continuum spectrum,  the $L^2$--norm of the
physical states is infinite. In the coordinate representation,
a function
$\psi_0(x)$ is annihilated by the constraint operator $p_y = -i 
\partial_y$,
but it clearly does not have finite norm in the original Hilbert space
with inner product given by $\int_{{\rm I\! R}^2} dx dy$. Similarly, in the
$p_y$-polarization the physical wave functions $\psi_0(x)
\delta(p_y)$ lead to the ill--defined square of a delta function. It
is obvious also from the form of the BRST charge $Q$ that,
in an appropriate
polarization of the wave functions, $\psi_0(x)$ is also BRST--closed.
The norm then contains the infinity obtained in the Dirac approach,
multiplied here by a zero from the (Berezin)
integration of the Grassmann
variables, and thus is ill--defined as well, as already mentioned in
the Introduction.

Let us now apply the \BM procedure to the model. We first have to pick
a gauge condition. This is
trivial in the present case, let us choose $y=0$ as the simplest
possibility.  Similarly, in the nonminimal sector we choose $\l=0$.
Following the recipe of \BM$\! ,$ one then has to construct two hermitian
operator sets, subject to some consistency conditions (cf., e.g.,
\cite{M3}). In the present case a possible choice of these two
sets is:
\be
D_{(1)} := \{(y,{\cal C}),(i\bar{{\cal C}},\pi)\}  , \quad
D_{(2)} := \{(\l,\bar{{\cal P}}),(i{\cal P},p_y)\}\; . 
\label{D}
\ee Each set of operators consists of so--called BRST--doublets, which
means that the second operator in each round bracket is --- up to a
possible prefactor of $i$ ensuring hermiticity --- the
BRST--transformed (graded commutator $[Q,\cdot ]$) of the first
operator in the respective round bracket. So each of the two sets
$D_{(i)}$ consists of four operators in the present example, which in
turn may be grouped into two doublets.  We remark here that the two
sets $D_{(i)}$, $i=1,2$, are not independent from one another. It is a
generic feature of the two sets that one of them contains the gauge
condition of the minimal sector and the constraint of the nonminimal
sector and vice versa for the other set; together with the doublet
structure this is the basic principle behind the consistency
conditions required for the sets.

Next, one has to decide for one of the two sets, say $D_{(2)}$, and to
determine its kernel, i.e.\ the simultaneous kernel of all four
operators of this set. Choosing a polarization such that all momenta
are represented by derivative operators except for $\pi$, which we
take as multiplication operator, this kernel is spanned by the
BRST--closed functions $\psi_0(x)$. These are the so--called auxiliary
states denoted by $|s\>_0$ in the Introduction. In the present
polarization the coincide with the physical wave functions of a Dirac
quantization.  
To obtain inner product states $|s\>$, the following (\BM)
procedure is applied: Multiply the respectively first entry of each
of the two doublets of the {\em other}\/ operator set, 
i.e., of  $D_{(1)}$ in our case, to obtain a gauge
fixing fermion $\Psi \equiv i y {\bC}$ and define $|s\> := \exp
[Q,\Psi]_+ \,|s\>_0$. Since $Q$ and $\Psi$ are hermitian, one finds
\footnote{ But cf. also the last paragraph in this section and the
comments to Eq.\ \re{master1}) in the Introduction.}
\be
\<s|s\> = {}_0\<s| \exp \( 2[Q,\Psi]_+\) | s\>_0 \, .  
\label{master}
\ee In explicit terms the above formula reads \ba \<s|s\> &\propto&
\int dx \, dy \, d \wt \pi \, d {\cal C}\, d \bar{\cal C} \,\, \psi_0^*(x) \,
\left[\exp \( -2i\wt \pi \, y \)
\(1+2 {\cal C} {\bar{\cal C}} \) \right] \; \psi_0(x) \nn \\
&\propto& 2\int dx \, dy \, \;\; \psi_0^*(x) \, \d(2y) \; \psi_0(x)
\,\, \equiv \int dx |\psi_0(x)|^2 \,\, .
\label{ss}
\ea
Here $\wt
\pi \equiv i\pi \in {\rm I\! R}$ as $\pi$ has to be quantized indefinitely
\cite{Pauli,M4}. 
This latter fact is also the reason for the
above phase in front of $|s\>_0$ to {\em add up}\/ in the inner
product rather than to drop out from it: The original inner product
for the indefinitely quantized variable $\pi$ is of the form $\< f | g
\> = \int d \wt \pi \, f^*(\pi^*) \, g(\pi)$, where the wave functions
are understood as functions of the spectrum of the hermitian operator
$\pi$, which is purely imaginary in this case.  In fact,
the ghost degree of freedom $\bar
{\cal C}$ has also to be quantized indefinitely; however, for a Grassmann
variable this makes no difference in the end.

So, in comparison to the Dirac procedure the upshot of the \BM inner
product is to get rid of the gauge group volume $\int dy = \infty$ by
effectively introducing an appropriate delta function for this
integral (cf.\ second line in Eq.\ \re{ss})), leaving the ordinary
Lebesgue measure for the single physical variable $x$. 

It is worthwhile to have a look at the same analysis in the momentum
representation of the constraint $p_y$. After switching to the
momentum representation (only for this variable for simplicity), the
kernel of the operator set $D_{(2)}$ again coincides with the Dirac
physical wave functions of the respective polarization (no additional
ghost terms occur, as is the case for other polarizations of the wave
functions): $\psi(x) \, \delta(p_y)$. By means of Eq.\ 
\re{master}) one finds now
 \ba \<s|s\>&\propto& \int dx\,dp_y\,d\wt\pi\,
\(\psi_0^*(x)\d(p_y) \) \exp(-2\wt \pi \frac{d}{dp_y})
\psi_0(x)\d(p_y)\nonumber\\& =& \int dx\,dp_y\,d\wt\pi\,
\psi_0^*(x)\d(p_y) \psi_0(x)\d(p_y - 2 \wt \pi)\propto\int
dx\,|\psi_0(x)|^2 \, \, \label{ss2} . \ea Thus, the meaningless square
of the delta--function in the original inner product for the Dirac
states becomes ``regularized'' through the \BM procedure by the point
splitting in the product of the $\delta$--functions (cf.\ second line
in Eq.\ \re{ss2})), again leading to a well--defined $L^2$ norm over
the physical variable.

Summarizing, we see that the \BM scheme indeed resolves
the aforementioned problem of the inner product for physical
states.
An essential ingredient of their procedure
is the choice of an appropriate gauge condition, which underlies the
construction of the operator sets \re{D}) and eventually specifies
the norm (\ref{master}). In realistic
gauge theories,  one often is plagued by the Gribov problem,
excluding the existence of global gauge conditions. Strictly speaking,
this excludes also the existence of two operator sets $D_{(i)}$
fulfilling all the requirements of \cite{M1}.
In the sequel we study, with  examples of simple gauge
theories {\em with}\/ a Gribov problem, whether the local \BM
procedure, which ignores subtleties of the gauge fixing, can
be extended to a global level.

Finally we remark that in the argumentation within this section we
in part remained quite formal, e.g., when speaking of hermiticity or
cohomology classes of various operators --- we never specified domains
of definition of them. In the following, however, 
we intend to take equation (\ref{master})
as a definition, analyzing its consequences with care and accuracy.

\section{An irreducible\ abelian\ model\ with\ Gribov obstruction: The
  $SO(2)$--model}

In this section we consider a two--dimensional model with gauge group
$SO(2)$. The gauge orbits are concentric circles (one--spheres), 
 generated by the constraint $G\equiv l=x_1p_2-x_2p_1\approx 0$
\cite{jackiw,lvp}. Obviously, this
is the angular momentum in two dimensions. The reduced phase space is
the half--plane ${\rm I\!R}_+\times {\rm I\! R}$ with the
identification $(0,p)\sim(0,-p)$, which is homeomorphic to a cone
\cite{lvp,ufn}. We draw attention to the {\em nontrivial topology}\/
of the gauge orbits as in contrast to the model of the previous section
where the gauge orbits are just parallel straight lines.
This fact gives rise to the non--Euclidean geometry of the reduced phase
space and will play a crucial role in the subsequent analysis.

The classical Hamiltonian of the model is simply
\be H=\frac{1}{2}{\bf
  p}^2+V({\bf x}^2)+\lambda l\;,
\ee
where $V({\bf x}^2)$ is some gauge invariant
potential and $\lambda$ a Lagrange multiplier enforcing the constraint
$l=0$. Here we shall again adopt the nonminimal BRST scheme and
treat $\lambda$ as a dynamical variable with conjugate momentum
$\pi$. The extended model has the further constraint $\pi\approx 0$,
generating orbits isomorphic to $\dR$. Then the nilpotent BRST charge
becomes $Q={\cal C}l+\pi\bar{\cal P}$.

The Gribov obstruction arises from the nontrivial topology of the
gauge orbits and is already obvious at this stage: There exists no
single--valued, globally regular function $\chi(x_1,x_2)$ such that
the gauge fixing curve $\chi=0$ intersects each gauge orbit precisely
once. In the \BM $\,$ procedure, one has to specify two
operator sets of the BRST doublets satisfying some consistency
conditions. Among these is a condition that essentially states that
the Faddeev--Popov determinant of the underlying gauge conditions has
to be {\em nonzero}\/.  This requirement {\em cannot} be met. However,
the deficiency may be localized to a single point on the gauge fixing
line in the configuration space, spanned by $(x_1,x_2)$, namely to the
origin $x_1=x_2=0$.  Let us ignore, for a moment, this seemingly small
deficiency and proceed with the \BM construction of a scalar product.
The two sets of operator doublets are chosen to be
\begin{eqnarray}
D_{(1)}&=&\{(x_2,x_1{\cal C}),(i\bar{\cal C},\pi)\}\ ,\\
  D_{(2)}&=&\{(i{\cal P},l),(\lambda,\bar{\cal
    P})\}\;.\nonumber
\end{eqnarray}
Now we determine the
kernel of the set $D_{(2)}$. In a convenient polarization it reads
$\<{\bf x},\pi,{\cal C},\bar{\cal C}|s\>_0=\psi_0({\bf x}^2)$. The
hermitian gauge fixing fermion is $\Psi=ix_2{\cal C}$ in this case.
With these ingredients we may now apply Eq.\ (\ref{master}) to obtain
the following inner product between two physical states:
\be
\<s|s^{\prime}\>\propto\int\limits_{\rm I\!
  R}dx_1\,x_1\psi_0^{\ast}(x_1^2)\psi_0^{\prime}(x_1^2)\equiv 0\;.
\label{neun}\ee
Thus, the \BM procedure does {\em not}\/ lead to a well--defined
physical scalar product here. This can also be verified for other 
polarizations of the wave function. 

In the particular polarization chosen here it is possible to obtain a
scalar product by some simple {\em additional}\/ manipulation, e.g.\
by restricting the range of integration to the positive axis or by
replacing the integration measure $x_1$ by $|x_1|$. However, this
would not be in the spirit of the \BM procedure: As outlined in Secs.\
1 and 2, the idea was to keep the {\em original}\/
inner product and to just select appropriate BRST--representatives in
order to yield well--defined amplitudes. The original measure was not
to be altered.

One could try to think of another gauge fixing fermion $\Psi$
in (\ref{master}) that would lead to a more appropriate measure
like $|x_1|$ or $x_1\theta(x_1)$ with $\theta(x_1)$ being the
characteristic function of the Gribov domain $x_1>0$. However,
it is not hard to  convince oneself that such a gauge fixing fermion
cannot be of the conventional form $\Psi = i\chi(x_1,x_2)\C$
for any smooth single--valued function $\chi$. So, the vanishing of the
FP determinant even at a single point appears to be quite an
obstacle to a naive global extension of the \BM procedure.
We shall see, however, that in the reducible case the situation
turns out to be better.

\section{A\ reducible\ \ nonabelian\ \ model\ with\ Gribov 
obstruction: The $SO(3)$--model}

In this section we apply the \BM procedure to a mechanical model 
with gauge group $SO(3)$ \cite{christ,lvp}.                                    
The new feature of this model, besides being nonabelian, is the
reducibility of the constraints generating the gauge orbits.  The
constraints $G_a$ are given by the three components of the angular
momentum $G_a \equiv l_a=\varepsilon_{ab}{}^cx^bp_c$. It is easy to
see that the $G_a$ can be combined nontrivially to zero: $x^aG_a\equiv
0$. The reducibility arises from the fact that the gauge orbits, which
are two--spheres, are not parallelizable. They do not admit one
globally nonvanishing vector field. In general, irreducible theories
can be turned easily into reducible ones by adding constraints that
are not independent of the original ones, but, as demonstrated already
by the above example, not necessarily vice versa. This is what we
called ``essential reducibility'' in the Introduction.

In the context of BRST--quantization, the reducibility of the
constraints is taken into account by the introduction of additional
ghost--of--ghost variables. The total number of variables in a
BRST--quantization blows up considerably with increasing rank of the
Lie algebra of the considered model, especially when one deals with
nonminimally extended models.  The dependences of the constraints can
be given by $Z_A^aG_a=0$ with phase space functions $Z_A^a(x,p),\;
a=1,\ldots ,d;\;A=1,\ldots ,r $.  In what follows the functions
$Z_A^a$, which exhaust all possible reducibilities, turn out to be
independent. So there will be no need for ghost--of--ghosts of higher
rank, unless explicitly stated otherwise as in the models of Sec.\
6.1 below. 

The $r$ sets of functions (vectors) $Z_A^a$ may always be multiplied by some
nonvanishing functions $f_A(x)$ without spoiling their characteristic
feature of specifying all independent reducibilities of the
constraints $G_a$.  We will see that, upon following the \BM
procedure, this arbitrariness in the choice of the functions $f_A$
will yield {\em different}\/ measures in the physical scalar products.
Moreover, only for a {\em subset}\/ of such functions $f_A$ or,
equivalently, for a subset of vectors $Z_A^a(x,p)$, the resulting
measure will make sense (i.e.\ will be physically acceptable). Still, also
in these cases the measure will depend explicitly on the specification
of those functions. However, the algebra of observables will then 
turn out to be modified accordingly so that the representations with
different choices of $f_A$ (from the allowed subset) turn out to be
equivalent (and in particular also unitarily equivalent to the
covariant Dirac result). 

In the present model we found one relation ($r=1$) between the
constraints $G_a$ with $Z_1^a \equiv x^a$ as a possible choice. In
this case the ``allowed subset'' of nonvanishing functions $f_1(x^a)$
will turn out to comprise the (nonvanishing) gauge invariant
functions of $x^a$, i.e.\ $f_1= f({\bf x}^2)$.\footnote{In fact the
  characteristic feature of the allowed subset is merely to have
  $|f_1|$ invariant under $x_1 \to - x_1$. We will come back to this
  issue.} In the following we will first restrict ourselves to this
``gauge invariant'' parameterization of the dependences, discussing
changes that occur when allowing for more general functions $f_1$ only
at the end.

In the $SO(3)$--model the BRST--operator $Q$ is given by
\begin{equation}
Q={\cal C}_0^aG_a+if({\bf x}^2){\cal C}_1^1x^a{\cal
    P}_{0a}-\frac{1}{2}\varepsilon_{ab}{}^c{\cal C}_0^a{\cal
    C}_0^b{\cal P}_{0c}+\pi_{0a}\bar{\cal P}_0^a+\pi_{11}\bar{\cal
    P}_1^1+\pi^1_{11}\bar{\cal P}_1^{11}\;,
\end{equation}
where indices $a,b,c$ run from $1$ to $3$, $\varepsilon_{ab}{}^c$ are the
structure constants of the Lie algebra $so(3)$ of 
the group $SO(3)$, and $f({\bf
  x}^2)$ is some nonvanishing function of ${\bf x}^2$. The first term
in the BRST--charge contains the constraints $G_a$, the second term
reflects their mutual dependence with $f({\bf x}^2)$ parameterizing the
arbitrariness mentioned above, the third term, cubic in
the ghosts, is standard for nonabelian groups \cite{marc}, and the last
three terms are a result of nonminimal extension ({\em including}\/
the ghost--of--ghost sector \cite{Batalin}). The full extended phase
space of the model consists of nine bosonic and eight fermionic pairs
of canonically conjugate variables. The bosonic pairs are $(x^a,p_a),
(\lambda^a,\pi_a),({\cal C}_1^1,{\cal P}_{11}),(\bar{\cal
  C}_{11},\bar{\cal P}_1^1)$ and $(\lambda^{11}_1,\pi_{11}^1)$, while
the fermionic pairs are $({\cal C}_0^a,{\cal P}_{0a}),(\bar{\cal
  C}_{0a},\bar{\cal P}_0^a),(\lambda_1^1,\pi_{11})$ and $(\bar{\cal
  C}_{11}^1,\bar{\cal P}_1^{11})$.\footnote{Here the notation conforms
    to the one used in \cite{Batalin}.} The number of
bosonic pairs exceeds the number of fermionic ones by one. So the model
possesses one bosonic physical degree of freedom.  The
unphysical sector has eight fermionic and eight
bosonic degrees of freedom.
In quantum theory both the bosonic and the
fermionic degrees of freedom of the unphysical sector have to be
quantized with half positive and half indefinite metric states.

Following the \BM procedure we construct two sets of
hermitian BRST doublets
\begin{eqnarray}
D_{(1)}&=&\{(0,0),(x^2,{\cal C}_0^1x^3-{\cal
C}_0^3x^1),(x^3,{\cal C}_0^2x^1-{\cal
C}_0^1x^2),(\lambda_1^{11},\bar{\cal P}_1^{11}),\\ & &({\cal
C}_0^1,i{\cal C}_0^3{\cal C}_0^2-
f({\bf x}^2){\cal C}_1^1x^1),(i\bar{\cal
C}_{0a},\pi_{0a}),(\bar{\cal
C}_{11},\pi_{11})\}\;,\nonumber\\
D_{(2)}&=&\{(0,0),(i{\cal
P}_{02},G_2-\varepsilon_{2b}{}^c{\cal C}_0^b{\cal P}_{0c}),(i{\cal
P}_{03},G_3-\varepsilon_{3b}{}^c{\cal C}_0^b{\cal P}_{0c}),
\nonumber\\ & &
({\cal P}_{11},if({\bf x}^2)x^a{\cal P}_{0a}),(i\lambda_1^1,\bar{\cal
P}_1^1),(\lambda_0^a,\bar{\cal P}_{0a}),(i\bar{\cal
C}_{11}^1,\pi_{11}^1)\}\;.\nonumber
\end{eqnarray}
Here a short comment about the admittedly somewhat strange doublet $(0,0)$ is
in place: The first three doublets of the set $D_{(1)}$ comprise
essentially three gauge fixing conditions $\chi^i$ together with their
BRST transforms $[Q,\chi^i]=iM^i_j{\cal C}^j,\;i,j=1,\ldots
,3$. Because of the reducibility of the model the FP matrix $M^i_j$
is at most of rank two. So it is possible and convenient to make the
trivial choice $\chi^1\equiv 0$ for the first of the gauge fixing
functions. (In other words, locally, i.e.\ up to some regions in phase
space of a lower dimension, only two of the three constraints
are essential and thus only two gauge conditions, $\chi^2=0$ and
$\chi^3=0$, are necessary). An analogous reasoning applies to the 
trivial doublets $(0,0)$ in the set $D_{(2)}$ above as well
as in the operator sets of the models to be discussed below.

The following steps are like those in Sec.\ 2.  First, we evaluate
the kernel of the operator set $D_{(2)}$.  As before it is given by
gauge invariant functions $\psi_0({\bf x}^2)$ as the physical
wave functions in the Dirac 
quantization . They are the auxiliary
states $|s\>_0$. The hermitian gauge fixing fermion $\Psi$ is bilinear
in those operators of the set $D_{(1)}$ which are not BRST invariant
(first entries of doublets).
A possible choice is \be \Psi=i\bar{\cal C}_{02}x^2+i\bar{\cal
  C}_{03}x^3+\bar{\cal C}_{11}{\cal C}_0^1+i\bar{\cal
  C}_{01}\lambda_1^{11}\,.\ee Now we can compute the norm of a
physical state by means of Eq.\ (\ref{master}). Again one has to keep
in mind that half of the unphysical variables must be quantized with
indefinite metric. After some calculation one finds \be \<s|s\>
\propto \int
dx_1\,\frac{|x_1|}{|f(x_1^2)|}\psi_0^{\ast}(x_1^2)\psi_0(x_1^2)\, .
\label{so31} 
\ee
In an intermediate step in deriving (\ref{so31}), we made use of the
formula \be \int\limits_{-\infty}^{\infty} dx \, dy \, \d(xy)x^2
\varphi(x)= \int\limits_{-\infty}^\infty dx \, |x| \varphi(x)\ ,
\label{test} \ee where $\varphi$ is some smooth (test) function. 
This relation may be obtained, e.g., by interpreting $\delta(z)$ as
the limit of an arbitrary delta sequence $\delta_n(z)$.  The above
integral equality is then understood as a limit of the sequence of the
integrals in the l.h.s.\  with $\delta(xy) \rightarrow \delta_n(xy)$.
The limit does not depend on a particular choice of the delta
sequence.  This may be proven by going over to new integration
variables $\widetilde x=xy, \widetilde y =x/y$ (after splitting the
integral into four integrals over regions with a definite sign of $x$
and $y$ to make the coordinate transformation well--defined) and by
making use of the characteristic properties of the delta sequence
$\delta_n(\widetilde x)$. Alternatively, the relation \re{test}) may
also be obtained by the means of \cite{GelfandI} (cf. chapter
III.4.5). 

In a final step we may now rewrite the r.h.s.\ of Eq.\ \re{so31})
identically by switching to a new integration variable $C:=(x^1)^2$
as follows:
\be  \int\limits_{\dR_+}\frac{dC}{|f(C)|} |\psi_0(C)|^2 \,\, .
\label{so32} \ee Note that $C$ may be interpreted also as the gauge
invariant Casimir polynomial $C \equiv {\bf x}^2$, expressed in the
gauge $x^2=x^3=0$ (we therefore use the same symbol). The result
\re{so32}) for the norm of a physical state $|s\>$ represented by
$\psi_0(C)$ is now a well--defined ($f$ was required to be
nonvanishing), physically sensible (states are represented by
functions of the only independent gauge invariant quantity
$C$; the ambiguity in $f$ will be discussed shortly) positive definite
inner product.

It is now in place to discuss changes that are induced by a more general
choice for the function $f_1$ in $Z_1^a$. Had we chosen $f_1$ as an
{\em arbitrary}\/ nonvanishing function of $x^a$, $f_1=f_1(x^a)$,  
all the steps leading to Eq.\ \re{so31}) may be 
repeated, we only need to replace $f({(x^1)}^2)$ by $f_1(x^1,0,0)$ in
this formula. Now we arrive at the curious conclusion that {\em only}\/ if
$|f_1|$ is invariant under a change of sign of $x_1$, we obtain a
physically acceptable inner product!\footnote{This is a necessary and
sufficient condition in the present ansatz for quantizing the model. 
The restriction to gauge invariant prefactors $f$ was just made 
for convenience, which would, in particular, simplify a similar
discussion, when starting with the operator set $D_{(2)}$ instead of
$D_{(1)}$ chosen for the present treatment.} Otherwise, with a
noninvariant function $f_1$, the {\em gauge equivalent}\/ positive and
negative half--axis of $x_1$ are weighted differently, which is
incompatible with the principle of gauge invariance. 

Restricting ourselves to the allowed subset of invariant functions
$f_1$ or $f$ as discussed above, there is still the apparent ambiguity
of the result (\ref{so32}) in choosing $f$, which may seem puzzling at
first sight.  To 
shed some light on this issue, we briefly illustrate the situation
with the example of the model discussed in Sec.\ 2, turning it
(artificially) into a reducible theory by counting the constraint
$p_y=0$ {\em twice}\/: $G_1=G_2=p_y$. This reducibility may be built in
by means of the relation $f(x)\,(G_1-G_2) = 0$ with some arbitrary
nonvanishing function $f$. Adapting the steps of Sec.\ 2 to the
reducible case, one finds that Eq.\ \re{ss}) becomes replaced by $\<
s| s\> \propto \int dx \, |\psi_0(x)|^2/|f(x)|$ --- in complete
analogy with \re{so31}).

The presence of the function $f$ in the measure poses no problem per
se; after all, we may absorb it by redefining the wave functions
$\psi_0(x)$: $\psi_0 \ra \psi := \psi_0/\sqrt{|f|}$.  Such an 
ambiguity of
the measure goes hand in hand with changes in the representation
of the momentum operator: If, as usual, $p\psi =-i\,
(d/dx) \, \psi$ for the standard measure  $dx$,
then it has to
take the form $-i\,|f|^{1/2} $ $ (d/dx)\, |f|^{-1/2}$ $ = -i\, d/dx +
i (\ln |f|)'(x)/2$ when applied to $\psi_0$. Moreover, it is only
this latter expression that allows for hermiticity of $p$ in the
space of square integrable functions with the
measure $dx/|f(x)|$. With this definition of $p$ both 
representations of the quantum theory become unitarily equivalent and
thus physical amplitudes are unaffected by the choice of $f$.

Now we return to the inner product defined by \re{so32}).  We will
show that also here the physical amplitudes are independent of the
choice of $f$ and coincide with those in the gauge invariant approach
of Dirac.\footnote{Recall that in this simple model the gauge group is
compact (and of finite volume $4\pi$) and thus the inner product of
the original Dirac quantum space remains well--defined also for
physical states.} As the $SO(3)$--model has one physical degree of freedom,
we have to find one further independent gauge invariant observable
beside the Casimir $C = {\bf x}^2$.  Restricting ourselves to an
observable that is at most linear in derivatives in the ${\bf
x}$--representation, the simplest hermitian choice is \be {\CO}:= \2
\( {\bf x} {\bf p} + {\bf p}{\bf x} \) \, .
\label{Q}
\ee
Its commutation relation with $C$ is $[C,\CO]=2i C$, forming an
affine algebra. Note that $\CO$ is an algebraically well--defined
object, while a gauge invariant canonical conjugate to $C$ does not
exist globally; such a conjugate operator would be $\CO/2C$, but $C$
may have zeros or, on the operator level, it is not invertible. 

To find the action of $\CO$ on the auxiliary state  $\psi_0(x)=
\<x|s\>_0$, we have to apply this
operator to $\<x|s\>$, on which the operator $\CO$ acts
according to the definition (\ref{Q}) where $p_j$ is
represented by $-i \6/\6x^j$ , and pull it through the
operator $\exp\([Q,\Psi]_+\)$, cf.\ Eq.\ \re{master}). Integrating out
all variables except for $x_1$ or $C$, respectively, a straightforward
calculation yields
\be \<s|\CO|s\> \propto
\int\limits_{\dR_+}\frac{dC}{|f(C)|} \psi_0^*(C) ( C \, P_C + P_C
\, C ) \psi_0(C) \,\, 
\label{Oso3}
\ee with
$P_C \equiv -i |f(C)|^{1/2}
\frac{\displaystyle d}{\displaystyle dC}|f(C)|^{-1/2}$.

Thus, we observe that, first, the action of
$\CO$ on $\psi_0$ depends on $f$.  Second, by construction 
$\CO$ still
has the correct commutation relations with $C$.  Moreover, upon an
appropriate choice of boundary conditions for the physical wave
functions $\psi_0(C)$, it is hermitian with respect to the effective
inner product of Eq.\ \re{so32}).  In complete analogy with the
reducible version of the model with a translational gauge symmetry,
one then easily establishes that all physical results are independent
of the choice of $f$. The nontrivial representation of $\CO$ for a
given function $f$ is essential in this context, however. Note also
that all gauge invariant observables of the original theory may be
expressed in terms of $C$ and $\CO$ and, up to usual operator ordering
ambiguities, can be represented as operators of the theory
defined on the positive real axis $C \in \dR_+$.

We finally want to specialize the result \re{so32}) to two particularly
nice choices for $f$. For $f:=1$ the measure in the Casimir variable
$C$ becomes trivial and the operator $P_C$ reduces simply  to $-i
d/dC$. The choice $f(C):= 1/\sqrt{C}$ with the simultaneous change of
variables to the ``radial'' coordinate $r := \sqrt{C}$ leads to the
measure $\int_{\dR_+} r^2 dr$, which one might favour as stemming from
a spherical reduction of $\int d^3x$, while the operator $\CO$ may be
shown to turn into $\CO = (r p_r + p_r r)/2$ where $p_r = -i
\frac{1}{r} \frac{d}{d r} r$ is the radial momentum operator, $p_r =
({\bf x} \cdot {\bf p} + {\bf p} \cdot {\bf x})/2r$. Up to an
irrelevant factor of $4\pi$, the latter measure is the one found in the 
Dirac quantization, thus proving unitary equivalence of the \BM
result \re{so32}) and the covariant Dirac result as promised.  

Despite the similarity of the Gribov problem in both the models
studied in this and the previous section, the topological obstruction
to the global extension of the \BM procedure appears to be not that
fatal in the present $SO(3)$--model as it was in the
$SO(2)$--model. As already remarked in the Introduction, much of the
motivation for studying further models in this paper is to
find possible reasons for why the \BM procedure fails for the
$SO(2)$--model while it works for the $SO(3)$--model. Moreover, in the 
latter case it worked only when  some restrictions were placed on the
functions 
characterizing the dependences of the constraints; we also want to
find the analogous restrictions in more general models where the \BM
procedure works despite the Gribov obstruction.

\section{Mechanical models with a semisimple gauge group in
  the adjoint representation}

In this section we want to 
extend the considerations of the previous section 
to arbitrary semisimple gauge groups. For this purpose we interpret 
the action of $SO(3)$ on the three--dimensional configuration space
$\dR^3$ as the {\em adjoint}\/ action of $SO(3)$ on its Lie algebra. 
To define the model we then only have to replace $SO(3)$ by a general
compact semisimple Lie group $G$.  
In a way these models are $(0+1)$--dimensional nonabelian Yang--Mills
theories, cf.\ \cite{ufn,lvsh89,sh89} for a definition of these models on the
Lagrangian level.  

For pedagogical reasons the analysis is carried through in the
detail for the $SU(3)$--model  first
(Sec.\ 5.1). The generalization to arbitrary $G$ is then
straightforward and contained in Sec.\ 5.2.

\subsection{The $SU(3)$--model}

Let the $3\times 3$ matrices $\tau_a,
\;a=1,\ldots ,8$, be the generators of the $su(3)$ Lie algebra satisfying
$[\tau_a,\tau_b]=if_{ab}{}^c \tau_c$. The generators $\tau_1$ and
$\tau_2$ are chosen to be diagonal.  They generate the Cartan
subalgebra of $su(3)$. Within our conventions the Cartan--Killing
metric is trivial, $g_{ab}={\rm tr}(\tau_a\tau_b)=\delta_{ab},$ 
and the totally symmetric (ad--)invariant tensor $d_{abc}$ is
defined by 
\be \{\tau_a,\tau_b\}=\frac{2}{3}\delta_{ab}{\rm \bf
  I}_3+d_{ab}{}^c\tau_c\;.
\ee
The configuration space coincides with the Lie algebra itself, and
the physical motion is subject to eight first--class constraints: 
$G_a
\equiv f_{ab}{}^cx^bp_c \approx 0$.  

The
constraints are not independent from one another and satisfy 
the relations
$Z_A^aG_a=0,\;A=1,2$, making the model reducible. The
functions $Z_A^a$ are chosen to be 
\be 
Z_1^a=f_1(C_1,C_2)x^a,\quad
Z_2^a=f_2(C_1,C_2)d^a{}_{bc}x^bx^c\;,
\label{Z}\ee 
where $f_1$ and $f_2$ are
arbitrary nonvanishing functions of the two independent
invariant (Casimir) polynomials 
\be C_1=\delta_{ab}x^ax^b\qquad
{\rm and}\qquad C_2=d_{abc}x^ax^bx^c \label{Casi}.\ee
Again at this point we could allow for
arbitrary nonvanishing functions $f_A$ on the configuration
space. However, only a subset of these, 
containing the gauge invariant functions chosen
above, will provide a reasonable inner product in the end. We will
discuss this issue further below. 

It is readily seen that $Z_1^a$ provides a dependence among the
$G_a$. For $Z_2^a$ this follows from the relation
$f_{(a|b}{}^cd_{c|de)}=0$, where $(a|\cdots|de)$ means symmetrization
with respect to the indices
$a,d,e$.  Alternatively, the relations $Z_A^aG_a=0$ may be inferred
from the ad--invariance of the Casimir polynomials $C_A$,
$0=\{G_a,C_A\}\equiv - f_{ab}{}^cx^b (\partial C_A/\partial x^c)$ and
the fact that $\partial C_1/\partial x^a= 2 x^a$ and $\partial
C_2/\partial x^a= 3d^a{}_{bc}x^bx^c$.

The existence of two relations amongst the eight gauge constraints
implies six dimensional gauge orbits, leaving two physical degrees of
freedom.  The BRST charge $Q$ has the same structure as for the
$SO(3)$--model: \be Q={\cal C}_0^al_a-\frac{1}{2}f_{ab}{}^c{\cal
  C}_0^a{\cal C}_0^b {\cal P}_{0c} +i{\cal C}_1^AZ_A^a{\cal
  P}_{0a}+\pi_{0a}\bar{\cal P}_0^a+\pi_{1A} \bar{\cal
  P}_1^A+\pi_{1A}^1\bar{\cal P}_1^{1A}\;.\label{Q1} \ee The extended
phase space of this model is similar to the one in the preceding
model, just with more variables. There are 42 pairs of canonically
conjugate variables now, only two of which represent physical degrees of
freedom.  The rest of the phase space comprises both 20 bosonic and
fermionic unphysical degrees of freedom.

Performing the \BM procedure along the lines already
explained above, one obtains for the
scalar product of two physical states after some tedious computation
\be \<s|s^{\prime}\>\propto \int\limits_{{\rm I\! R}^2}dx_1\,dx_2\,
\frac{|x_1|}{|f_1(u,v)|}\frac{|3x_2^2-x_1^2|}{|f_2(u,v)|}\psi_0^{\ast}
(u,v)\psi_0^{\prime}(u,v)\;.
\label{su3}
\ee
 
 Here we used the abbreviations $u\equiv x_1^2 + x_2^2$ and 
$v\equiv x_2 \, (3x_1^2 -x_2^2)$, which respectively equal the
Casimirs $C_1$ and $C_2$ in the gauge $x_i=0,\, i=3,\ldots ,8$  chosen
to construct the gauge fixing fermion.

Let us remark first of all that the appearance of absolute value signs
around $x_1$ and $3x_2^2-x_1^2$ seems quite noteworthy to
us. Irrespective of the fact that anyway a scalar product necessarily has to
be positive, without these absolute value signs the inner product
would vanish identically! The reason is that
the Casimir functions $u$ and $v$ exhibit some residual gauge invariance,
known as the Weyl group $W$ (more on this below). E.g.\ they are obviously
invariant under $x_1 \ra -x_1$. 

A similar situation was encountered in the previous section: In
contrast to the $so(2)$--model, in the $so(3)$ case the $x_1$ in the
measure appeared as absolute value, ensuring nonvanishing of the inner
product. The symmetry $x_1 \ra -x_1$ related precisely those points on
the gauge fixing surface $x_2=0=x_3$ which were still gauge equivalent.

We are thus led to study the residual gauge invariance in the
$(x_1,x_2)$--plane in the gauge $x_i=0,\, i=3,\ldots ,8$ \cite{ufn}. 
As illustrated
in the left hand side of Fig.\ 1 a generic point has five gauge
equivalent ``Gribov copies''. The six gauge equivalent points may be
related to one another by (multiple) reflections with respect to the
lines $x_1=0$ and $x_2=x_1/\sqrt{3}$. This $Z_6$ is known as the
Weyl group $W$ of $su(3)$. Due to the absolute value signs the measure
is invariant under the full group $W$ and the inner product does not 
vanish due to this symmetry.

The analogy with $so(3)$ goes even further: Similarly to the
transition from Eq.\ \re{so31}) to Eq.\ \re{so32}), also the right
hand side of Eq.\ \re{su3}) can be expressed as an integral over the
Casimirs only, namely as \be \int\limits_{{{\rm Im} \,
    K^+}}\frac{dC_1}{|f_1(C_1,C_2)|}\,
\frac{dC_2}{|f_2(C_1,C_2)|}\,\psi_0^{\ast}
(C_1,C_2)\psi_0^{\prime}(C_1,C_2)\,.\label{RS}\ee Here the Casimir polynomials
$C_1$ and $C_2$ first arise in the change of variables from 
$(x_1,x_2)$ to $(C_1:=u(x^1,x^2),C_2:=v(x^1,x^2))$, but may also be
identified with the 
Casimirs \re{Casi}) of the original gauge invariant formulation of the theory.
Due to the Weyl invariance of the
functions $u$ and $v$, the map to the new coordinates is not
bijective: Each of the six modular domains, one representative of
which we denote by $K^+$, is mapped to one and the same region in the
$(u,v)$--plane, ${\rm Im}\, K^+$ (cf.\ hatched regions in Fig. 1).

\vspace{1cm}

{

\unitlength=0.6000000pt
\begin{picture}(550.00,250.00)(0.00,0.00)
\put(125.00,125.00){\line(-1,-2){10.00}}
\put(97.00,80.00){\line(12,19){12.00}}
\put(69.00,43.00){\line(13,18){13.00}}
\put(44.00,8.00){\line(12,17){12.00}}
\put(186.00,228.00){\line(10,17){10.00}}
\put(156.00,176.00){\line(-13,-21){13.00}}
\put(177.00,209.00){\line(-7,-11){14.00}}
\put(125.00,125.00){\line(13,23){13.00}}
\put(544.00,115.00){\makebox(0.00,0.00)[tl]{$v$}}
\put(434.00,248.00){\makebox(0.00,0.00)[l]{$u$}}
\put(284.00,217.00){\line(-10,-7){10.00}}
\put(285.00,218.00){\line(-13,5){13.00}}
\qbezier(180.00,173.00)(281.00,253.00)(461.00,200.00)
\put(461.00,200.00){\circle*{6.00}}
\put(180.00,173.00){\circle*{6.00}}
\put(196.37,139.81){\circle*{6}}
\put(101.84,55.79){\circle*{6}}
\put(144.11,55.77){\circle*{6}}
\put(50.92,138.57){\circle*{6}}
\put(76.49,179.4){\circle*{6}}
\put(300.00,225.00){\line(1,0){250.00}}
\put(323.00,205.00){\line(1,0){205.00}}
\put(390.00,185.00){\line(1,0){69.00}}
\put(412.00,165.00){\line(1,0){26.00}}
\put(421.00,145.00){\line(1,0){8.00}}
\qbezier(425.00,125.00)(418.00,205.00)(300.00,205.00)
\qbezier(425.00,125.00)(432.00,205.00)(550.00,205.00)
\put(425.00,0.00){\vector(0,1){250.00}}
\put(300.00,125.00){\vector(1,0){250.00}}
\put(134.00,249.00){\makebox(0.00,0.00)[l]{$x_2$}}
\put(241.00,119.00){\makebox(0.00,0.00)[tl]{$x_1$}}
\put(125.00,138.00){\line(11,-7){11.00}}
\put(125.00,158.00){\line(29,-17){29.00}}
\put(125.00,178.00){\line(5,-3){50.00}}
\put(125.00,198.00){\line(22,-13){66.00}}
\put(125.00,218.00){\line(5,-3){84.00}}
\put(125.00,239.00){\line(27,-16){103.00}}
\put(5.00,185.00){\line(2,-1){228.00}}
\put(125.00,125.00){\line(-2,1){120.00}}
\put(125.00,125.00){\line(2,1){116.00}}
\put(125.00,125.00){\line(-2,-1){121.00}}
\put(125.00,0.00){\vector(0,1){250.00}}
\put(0.00,125.00){\vector(1,0){250.00}}
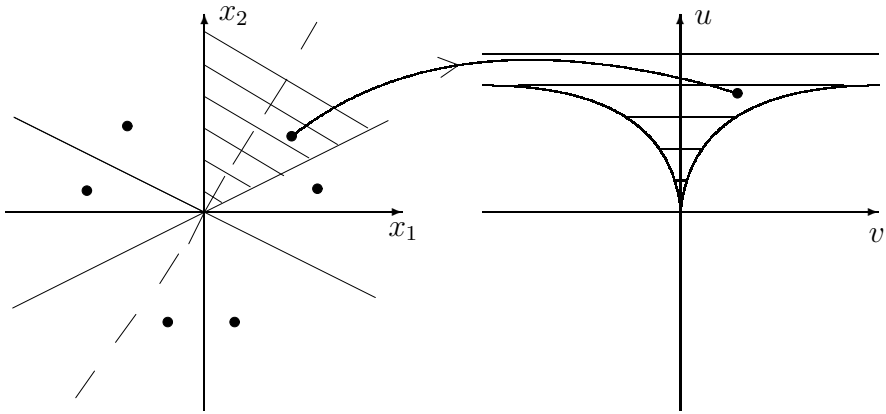
\put(0.00,-90){\parbox[tl]{335pt}{{\bf Fig.1}: {\em The left hand side
of the
picture shows  the residual gauge freedom in the case of $su(3)$:
The six dots are gauge equivalent. Upon the transition to Casimir
coordinates they are all mapped to the one point on the right hand
side of the picture.}}}
\end{picture}}

\vspace{3.5cm}

We remark that the functional determinant
of the map from $(x_1,x_2)$ to $(u,v)$, or ($C_1,C_2$), includes a
factor of $(1/6)$, which cancels precisely against the degree of the
map (the number of Gribov copies). As we will clarify in subsequent
sections (cf.\ in particular Sec.\ 7), this feature is rather
accidental and {\em not}\/ characteristic for the \BM procedure. By a
different normalization of the Casimirs there is no factor of  $(1/6)$
anymore and only a multiplicative factor equal to the number of 
Gribov copies remains. If this factor diverges, the \BM inner
product diverges as well, as will become most transparent in studying
the helix model in Sec.\ 7 below.

As for the $SO(3)$--model, one can supplement the operators
$C_{1,2}$ by two additional observables
$\CO_1 = p_a x_a$ and $\CO_2 = d_{abc} p_a x_b x_c$ (hermitized
appropriately) and prove that the physical amplitudes do not depend
on the choice of the functions $f_{1}$ and $f_{2}$. 

Also now we are in the position to analyze modifications that occur
when replacing the functions $f_{1},f_{2}$ of the Casimir coordinates in
Eq.\ \re{Z}) by arbitrary nonvanishing functions of the Lie algebra
coordinates $x^a$. As before, the transition from Eq.\
\re{su3}), where now the arguments of $f_1$ and $f_2$ are replaced by
$(x^1,x^2,0,0,0,0,0,0)$, to Eq.\ \re{RS}) will no more be possible in
general. A necessary condition for this transition is that $f_1$ and $f_2$
are invariant with respect to  the residual gauge freedom $W$ left by
our choice $x^3=\ldots =x^8=0$. Otherwise gauge equivalent sectors in
the $(x^1,x^2)$--plane would receive different weights, yielding an
unacceptable inner product.

Still the above condition  in the freedom of
choosing $Z_A^a$ depends on the gauge. The necessary and sufficient
gauge independent condition on the functions $f_1$ and $f_2$ is that they
are (nonvanishing) functions of the Casimir polynomials $C_1,C_2$ only,
as in our original ansatz in Eq.\ \re{Z}). Reformulating this
condition directly for the vectors  $Z_A^a$ (instead of just for the
functions $f_{1},f_{2}$ defined through  Eq.\ \re{Z})), one obtains that
the most general form of the these vectors that produces a
well--defined and acceptable inner product within the \BM version
presented here is:  \be 
Z_A^a = f_A(C_B) \, \frac{\6 F_A(C_B)}{\6 x^a} \, \, ,\label{gen} \ee 
where the $f_A$ are nonvanishing functions of the Casimir
polynomials in Eq.\ \re{Casi}) and det$(\6 F_A(C_B)/\6 x^a)\neq 0$. 
Actually, this parameterization of the  $Z_A^a$ results from our
previous ansatz \re{Z}) by a change of coordinates in the space of
Casimirs from $C_A$ to $F_A(C_B)$.

\subsection{Generalization to arbitrary semisimple groups}

The $SU(3)$--model studied in detail in the previous
subsection 
can be generalized to models of point particles transforming in the
adjoint 
representation of arbitrary semisimple Lie algebras $g$. 
We will see that increasing the number of physical degrees of
freedom does not affect the conclusion of the previous section:
the (appropriately refined) \BM inner product is well--defined for the 
reducible case despite the presence of a Gribov obstruction.

Now the variable
$x=x^a\tau_a,\;a=1,\ldots, d={\rm dim}\,g$, takes values in a 
semisimple Lie
algebra $g$. Here $\tau_a$ denotes a basis in $g$,
where we choose the convention that the 
first  $r={\rm rank}\,g$ generators span 
a Cartan subalgebra $H$ of $g$: $\tau_a = (\tau_{\mu},\tau_i)$, 
$\mu=1,\ldots, r$; $i=r+1, \ldots ,d$. 
The variable $x$ transforms according to the adjoint
action of the respective Lie group $G$. This action is generated by
the $d$ first--class constraints $G_a=f_{ab}{}^cx^bp_c$ with
$f_{ab}{}^c$ being the structure constants of $g$ and $p_a$ the
momenta canonically conjugate to $x^a$.

The constraints
fulfill $r=\dim H$ independent relations $Z_A^aG_a=0,\;A=1,\ldots,r$. 
The functions $Z_A^a$ read
\begin{equation}
Z_A^a=f_A \, \, C_A{}^a{}_{b_2\ldots
    b_{d(A)}} \, x^{b_2}\ldots x^{b_{d(A)}}\;. 
\label{x1x} 
\end{equation}
Here $C_{A,a_1\ldots a_{d(A)}}$ denote $r$ ad--invariant, symmetric,
irreducible tensors of rank $d(A)$ on the Lie algebra and the $f_A$ are
$r$ arbitrary nonvanishing functions of the Casimir polynomials
\be C_{A}=C_{A;a_1\ldots a_{d(A)}}x^{a_1} \ldots
x^{a_{d(A)}}.\label{Casallg}\ee  
For every semisimple group there is a polynomial of second order,
$d(1) =2$. For groups of rank 2, the degree of the second
invariant polynomial is $d(2) = 3,4$ and $6$ for $SU(3), SO(4)$
and $G_2$, respectively. The degrees of the Casimir polynomials
for groups of higher ranks can be found, e.g., in \cite{zhel}. 
We remark also that there is
no sum over the index $A$ in the right hand side of Eq.\ 
\re{x1x}). As in the previous subsection the relations
$Z_{A}^aG_a=0$ follow directly from the ad--invariance
of the symmetric tensors $C_{A;a_1 \ldots a_{d(A)}}$.
Our choice \re{x1x}) of the $Z_A^a$ is obtained from the general
ansatz of Eq.\ \re{gen}) by setting $F_A:=C_A$ and the restriction of
the $f_A$ to depend only on the $C_A$ is justified by the same
reasoning as in the $SO(3)$-- and the $SU(3)$--model.

The BRST--charge $Q$ is given by Eq.\ \re{Q1}), with the indices
running over the appropriate ranges now. Performing the \BM procedure
in the gauge $x^i=0,\;i=r+1,\ldots,d$, one obtains for the inner
product of two states
\begin{eqnarray}
\< s|s^{\prime}\>&\propto&\!\!\int\limits_{{\rm
      H}\times{\rm I\! R}^r}\!\! dx^{\mu}\,d\bar{\cal
    C}_{1\mu}\,\psi^{\ast}(C_1({\bf x}),\ldots,C_r({\bf x}))
  \left(\prod_{\alpha > 0}{\bf
      \alpha}\cdot{\bf x}\right)^2\times \label{ssx}\\
  &\times&\prod_{A=1}^r\delta(\bar{\cal
    C}_{1\mu}f_AC_A{}^{\mu}{}_{\nu_2\ldots\nu_{d(A)}}x^{\nu_2}\ldots
  x^{\nu_{d(A)}} ) \, \psi'(C_1({\bf x}),\ldots,C_r({\bf
    x}))\;.\nonumber
\end{eqnarray} 
Here ${\alpha}>0$
are positive roots of $g$.
They entered the calculation through the structure constants present in
the BRST charge,
which have the form
$f_{\m \a}{}^a = \delta^a_{\bf \a} \, \alpha^\m$, $\m = 1, ..., r$, 
if we assume  $\t^a = (\t^\m,\t^{\bf
  \a})$ to be the Cartan--Weyl basis \cite{zhel,hel}. 
However, the result does not depend on the choice of the basis.
The quantity
$C_{A;\mu_1 \ldots \mu_{d(A)}}$ is the pullback of the respective
Casimir tensor $C_{A;a_1 \ldots a_{d(A)}}$ under the embedding of the
chosen Cartan subalgebra $H$ (with coordinates $x^\mu$) 
into the Lie algebra $g$. Note
that as the Casimirs are ad--invariant, the tensors on $H$ are
independent of the embedding of $H$ into $g$, since any two 
Cartan subalgebras within $g$ are
related to one other by an adjoint transformation.  Also
Eq.\ \re{ssx}) is independent of the specific choice of the
ad--invariant tensors on $g$: A redefinition $C_{A;a_1\ldots
  a_{d(A)}}$ by $C_{A;a_1\ldots a_{d(A)}}+C_{B;(b_1\ldots
  b_{d(B)}}C_{C;c_1\ldots c_{d(C)})}$, where $d(A)=d(B)+d(C)$ and the
bracket indicates symmetrization over the smaller case indices, is
easily seen to have no effect.

With an appropriate normalization of the structure constants
and irreducible invariant symmetrical tensors, the following
relation holds 
\be 
  \prod_{\alpha > 0}({\bf \alpha}\cdot{\bf x})=
\det \( C_{A;\mu
    \mu_2 \ldots \mu_{d(A)}} x^{\mu_2}\ldots x^{\mu_{d(A)}} \)\,\, ,
\label{x}\ee
where the determinant is taken with respect to the two free indices
$A$ and $\m$, both of which range from one to $r$. 
We now substitute \re{x}) into (\ref{ssx}), make use of 
the multidimensional generalization of Eq.\ \re{test}),
\be \int d^rx
d^ry \, \prod_m \delta(a_{mn}(x) \, y_n) \, \left(\det a_{mn}(x)\)^2
  \varphi(x)= \int d^rx \left|\det a_{mn}(x)\right| \varphi(x) \ ,
\ee 
and then change the integration variables 
from $x^\m$ to $C_A$. This yields the
generalization 
\be 
\int\limits_{{\rm Im}\, K^+}
  \prod_{A=1}^r \frac{dC_A}{|f_A(C_1,\ldots, C_r)|} \;
  \psi^{\ast}(C_1,\ldots,C_r) \, \psi'(C_1,\ldots,C_r) 
\label{1}\ee 
of formula \re{RS})  for the right hand side of Eq.\ \re{ssx}).

Like in the $SO(3)$-- and $SU(3)$--case, imposing
$x^{r+1}=\ldots=x^d=0$ does not fix the gauge completely, but leaves
some discrete residual gauge freedom. In the context of Lie algebras
the above ``gauge fixing'' corresponds to a projection of the Lie
algebra to some representative  of the  respective Cartan subalgebra $H$
while the residual gauge freedom  is identified with the  Weyl group
$W$ \cite{lvsh89,ufn}. The Weyl group
consists of elements that are obtained
by all inequivalent compositions of reflections
in the hyperplanes orthogonal to simple roots of the
Lie algebra. A modular domain of $W$ on $H$ is called Weyl chamber
$K^+=H/W$.  
A possible representative of $K^+$ is 
$K^+=\{{\bf x}\in H|({\bf \alpha\cdot  x}) >0\quad \forall
\alpha>0\}$. For the special case of $SU(3)$ $W$ is generated by
$\hat{s}_1:\, (x_1,x_2)\rightarrow (-x_1,x_2)$ and 
$\hat{s}_2:\, (x_1,x_2)\rightarrow
(\frac 12[\sqrt{3}x_2 -x_1],\frac 12[-x_2-\sqrt{3}x_1])$ and $K_+$ may
be identified with a sector of angle $\pi/6$ in the two--plane,
cf.\ Fig.\ 1 of the previous subsection.

The reduction of the integration domain from the Cartan subalgebra $H$
to the Weyl chamber $K^+$, performed implicitly as one of the steps in
bringing \re{x}) into the form \re{1}), is possible\footnote{Besides
  the fact that the dependences were chosen in accordance with Eq.\
\re{gen}) certainly, cf.\ our
  foregoing discussion on this issue in the cases of $SO(3)$ and
  $SU(3)$.} since the number $N_W$ of modular domains or of Gribov
copies is finite.  In fact, as Fig.\ 1 illustrates for the case of
$SU(3)$, in this reduction the number $N_W$, which equals six in the
particular case of Fig.\ 1 but in general may be identified with 
$d-r$, $d \equiv {\rm dim}\, G$, appears as a multiplicative factor to
the inner product. Given our normalization of the latter polynomials
in Eq.\ \re{Casallg}), this number drops out from the final result
\re{1}) due to an exact cancelation\footnote{Use $N_W=\prod_{A=1}^r
  d(A)$.} with the Jacobian of the map
$x^{\mu}\rightarrow C_A(x^{\mu})$, performed in a {\em subsequent}\/
step. However, if the number of Gribov copies were infinite, such a
subsequent step would be impossible. Indeed, in Sec.\ 7 we will verify
explicitly by means of an example with an infinite number of Gribov
copies that in such a case the \BM inner product becomes divergent.

Concluding, we observe that, up to this stage, all the results
obtained from the \BM method coincide with those obtained from the
Faddeev--Popov (FP)
method (using the respectively same gauge conditions and the FP
determinant being defined as mentioned in the first footnote in the
Introduction). Indeed, for the $SO(2)$--model $\det \{x^2,l\} \equiv
x^1$ in coincidence with the 
measure found in formula \re{neun}). For the subsequent models, on the
other hand, it is not difficult to convince oneself that the
FP determinant is $\left(\prod_{\alpha
    > 0}{\bf \alpha}\cdot{\bf x}\right)^2$, evaluated in the gauge
chosen. This is nonnegative (cf.\ also \cite{lvsh89,sh89} for details)
and yields an inner product that coincides  with the one found in Eq.\ 
\re{1}) upon an appropriate choice of the
functions $f_A$. 

Thus the question arises, whether the \BM procedure works only in
those cases where the FP method does (despite a Gribov obstruction).
The models considered in the following will show that this is not the
case. There are also theories in which the \BM method works despite
the failure of the FP method.

\section{Models in the fundamental representation}

For the reducible models we have studied so far, the \BM procedure has
provided us with a well--defined inner product for physical states in
the BRST formalism, even in the presence of a Gribov obstruction and
for any finite number of physical degrees of freedom. As just mentioned,
these models exhibited the specific feature that the FP determinant
is nonnegative in the gauge used to construct the inner product
measure. Now we are going to demonstrate in Sec.\ 6.1 that this latter
feature is {\em not}\/ crucial for the existence of the global
extension of the \BM inner product for  reducible gauge models.
However, when the reducibility is removed, as will be done in Sec.\ 6.2 by
adding more degrees of freedom while keeping the gauge group fixed,
the \BM inner product becomes ill--defined due the Gribov topological
obstruction.

\subsection{$SO(N)$--model in the fundamental representation}

Here we study a point particle model with gauge group
$SO(4)$ in the fundamental representation and
its generalization to $SO(N)$. We have in mind to get 
a better understanding of the fact that the \BM procedure
yields an ill--defined inner product in the case $N=2$
and a well--defined one for $N=3$. {}From simple spherical
reduction, performed in the Dirac quantization after restriction to
(rotationally invariant) physical states, one would expect a measure
$r^{N-1},\; r^2:={\bf x}^2$. 
Unfortunately, this measure cannot be obtained by naive application of
the FP method. In the gauge $x^2=\ldots =x^N=0$ the FP determinant is
$(x^1)^{N-1}$. For even $N$ this is not positive definite on ${\rm I\!
R}$ and
leads to a vanishing inner product for gauge invariant wave functions.

We begin with the discussion of the
$SO(4)$--model which contains all essential features of
the general one with gauge group $SO(N)$.

Let the motion of a point particle in the configuration space
${\rm I\! R}^4$ be subject to the constraints $G_a={\cal
O}_a{}^i{}_jp_ix^j=0,\;i,j=1,\ldots ,4;\;a=
1,\ldots ,6=:\Gamma_0$ which are the angular momentum  
components in the eight--dimensional phase space.\footnote{$\Gamma_0$
as well as the subsequent $\Gamma$s and $\g$s are introduced for later
convenience when we generalize to $SO(N)$} ${\cal
O}_a{}^i{}_j$ form a basis of real antisymmetric $4\times 
4$-matrices.
The constraints are not independent, but
fulfill four relations $Z_A^aG_a=0,\;A=1,\ldots ,4=:\Gamma_1$.
The four 6-vectors $Z_A^a$ are   chosen to be linear in the
configuration space variables $x^i$. It is not hard to see that such a
choice is always possible. Certainly, again it would be  possible  to
multiply the vectors $Z_A^a$ by 
nonvanishing gauge invariant functions $f_A(r)$; for simplicity  they
are set to one in the following. The linearity of the functions
$Z_A^a$  ensures that they  are defined on the whole configuration
space, but, on the 
other hand, has the consequence, that the $Z_A^a$ are not independent: they 
combine to zero via a relation ${\cal Z}_1^AZ_A^a=0$, where also
${\cal Z}_1^A$ may be chosen linear in the $x^i$. It is  
easy to see that we have locally $6-4+1=3=:\gamma_0$ 
independent constraints, $4-1=3=:\gamma_1$ independent 
relations between the constraints, and $1=:\Gamma_2=\gamma_2$
independent relation of second stage. Thus, in contrast
to the other reducible gauge models discussed so far, 
the ghost--of--ghosts of higher rank must be introduced
in order to describe the dependence of the 
functions $Z_A^a$, called also null--eigenvectors of
the constraints \cite{Batalin}.

Following the general procedure proposed in \cite{Batalin},
we obtain the nonminimal BRST--charge
\begin{eqnarray}
Q&=&{\cal C}_0^aG_a-\frac{1}{2}f_{ab}{}^c{\cal C}_0^a
{\cal C}_0^b{\cal P}_{0c}+i{\cal C}_1^AZ_A^a{\cal P}_{0a}+
{\cal C}_2^1{\cal Z}_1^A{\cal P}_{1A}+\\& &+\pi_{0a}\bar{\cal P}_0^a
+\pi_{1A}\bar{\cal P}_1^A+\pi_{21}\bar{\cal P}_2^1+\pi^1_{1A}
\bar{\cal P}_1^{1A}+\pi^1_{21}\bar{\cal P}_2^{11}+\pi^2_{21}
\bar{\cal P}_2^{21}\;.\nonumber
\label{f}\end{eqnarray}
The first line of \re{f}) is the standard expression for a reducible
gauge 
model of rank two (i.e., with two stages of reducibility) in 
the minimal BRST approach. The second line contains the nonminimal
sector of $Q$. According to \cite{Batalin} we have several canonical
pairs of unphysical Lagrange multiplier, antighost, and extraghost
variables: The 
fermionic pairs $(\bar{\cal C}_{0a},\bar{\cal
P}_0^a)$, $(\lambda_1^A,\pi_{1A})$, $(\bar{\cal C}_{21},\bar{\cal
P}_2^1)$, $(\bar{\cal C}_{1A}^1,\bar{\cal
P}_1^{1A})$, $(\lambda_2^{11},\pi_{21}^1),$ and $(\bar{\cal
C}_{21}^2,\bar{\cal P}_2^{21})$, together with the bosonic pairs
$(\lambda_0^a,\pi_{0a})$, $(\bar{\cal C}_{1A},\bar{\cal
P}_1^A)$, $(\lambda_2^1,\pi_{21})$, $(\lambda_1^{1A},\pi_{1A}^1)$
, $(\bar{\cal C}_{21}^1,\bar{\cal P}_2^{11})$, and
$(\lambda_2^{21},\pi_{21}^2)$.

After choosing two consistent and convenient sets
of hermitian operator doublets, the \BM procedure
is  straightforward but tedious. The auxiliary states
come out to be gauge invariant states independent of
the ghost degrees of freedom, i.e., $\psi_0 =\psi_0({\bf x}^2)$.
For the gauge $x^2=x^3=x^4=0$, one obtains 
the inner product as an integral over the real $x$-axis of two wave functions
which depend on one variable $x^2\equiv (x^1)^2$. 
The resulting measure may be  constructed along the 
following lines: for every locally independent constraint take a factor
of $x$, for every locally independent relation between the constraints
a factor of $|x|^{-1}$ (this stems from a $\delta$--function mechanism
similar to the one in the previous sections), while one independent
relation between
the $Z_A^a$ gives rise to 
another factor of $x$. So the measure is $x^3\cdot|x|^{-3}
\cdot x =|x|$ which is positive definite on the entire real axis! 

Let us now see, how this procedure can be generalized to 
the group $SO(N)$.
Here we have $\Gamma_0={N \choose 2}$
constraints  
$G_{a_0}={\cal O}_{a_0}{}^i{}_jp_ix^j=0;\;i,j=1,\ldots ,N$. These
fulfill $\Gamma_1={N \choose 3}$ relations  
$Z_{a_1}^{a_0}G_{a_0}=0$.
The null--eigenvectors $Z_{a_1}^{a_0}$ of the constraints are not
independent and possess null--eigenvectors of their own, which
may also be linearly dependent, etc. The procedure goes on
up to the $(N-2)$-nd stage where there are no more linearly dependent
null--eigenvectors \cite{Batalin}.
On the $i$th stage of reducibility we have $\Gamma_i={N \choose i+2}$ relations
$Z_{a_i}^{a_{i-1}}Z_{a_{i-1}}^{a_{i-2}}=0$. All the $Z$s may be chosen
to be linear in the configuration space
variables $x^i$. $\Gamma_i$ is the smallest
possible number for which the $Z_{a_i}^{a_{i-1}}$ are well--defined on
the whole configuration space. 
Out of the $\Gamma_i$ 
relations $\gamma_i=\sum_{k=i+2}^N (-1)^{i+k}{N \choose k}={N-1
\choose i+1}$ are 
locally independent. Especially, $\gamma_0={N-1 \choose 1}=N-1$
constraints are locally
independent and so we have $N-(N-1)=1$ physical degree of freedom, which 
may be identified with the radial coordinate.

Now a generalization of the \BM inner product measure of
the $SO(4)$--model is straightforward.
For every locally independent relation of even stage (these include the
constraints $G_a$) the measure is provided with the factor $x$ and for
every locally independent relation of odd stage we have a factor of
$|x|^{-1}$. So one obtains for the measure
\begin{equation}\frac{x^{{N-1 \choose 1}+{N-1 \choose
3}+\ldots}}{|x|^{{N-1 \choose 2}+{N-1 \choose 4} 
+\ldots}}=\frac{x^{2^{N-2}}}{|x|^{2^{N-2}-1}}=|x|\; 
,\qquad\forall N>2\;.
\label{d}\end{equation}
More generally, multiplying the linear sets $Z_{a_i}^{a_{i-1}}$ by
nonvanishing gauge invariant (i.e.\ rotationally invariant)
functions, the  measure always takes the form $|f(x^2)x|$ (provided
$N>2$). Any (nonvanishing) function $f$ may be obtained in this way
leading to unitarily equivalent quantum theories (cf.\ foregoing sections).

We see that the \BM procedure leads for all $N>2$ to a well--defined
inner product. It does so  not only for odd $N$ like the FP method but
also for even $N$ where the latter failed. Moreover, the Dirac result
$r^{N-1}$ for the measure is reproduced upon the choice
$f(x^2)=|x^{N-2}|$.

Concerning our question why  \BM works in the $SO(3)$--model but
not in the $SO(2)$--model we are led to the following conclusion:
The failure of the \BM approach in the $SO(2)$--model was not due to
the inherent failure of the FP method for all $SO(N)$--models with
even $N$ in the
fundamental representation. The reason for the failure is assumed to
be the combination of the Gribov topological obstruction and the
irreducibility of the constraint(s). We will study this question in
the next subsection where the essential\footnote{cf. the discussion in the
Introduction} reducibility  of the constraints in the  $SO(3)$--model
is lost by increasing the number of physical degrees of freedom.

\subsection{Yang--Mills quantum mechanics}

In \cite{M1} Batalin and Marnelius have shown that their approach
is equivalent to the Faddeev--Popov procedure for models with
irreducible constraint algebras. As an example we have considered 
a particle
on the plane with the translational gauge symmetry 
and the $SO(2)$--model. In the presence of the  Gribov obstruction,
the Faddeev--Popov procedure
suffers from nonpositivity of the FP determinant and the subsequent
vanishing of the inner product for the conventional gauge fixing
fermion. In models with a reducible constraint algebra the
\BM procedure provided us with a mechanism to obtain a positive measure in
the inner product, which enabled us, under the assumption of a finite
number of Gribov copies, to construct a well--defined scalar
product. The transition from a reducible model to an irreducible one
can be made by adding more degrees of freedom subject to {\em simultaneous}\/
gauge transformations, while keeping the gauge group fixed.  
In doing so, we observe that the positivity of the measure
is lost, thus leading to an ill--defined inner product when
the Gribov topological obstruction is present.
 
We illustrate the statement by means of the example of Yang--Mills
mechanics. The model is obtained from the four--dimensional Yang--Mills field
theory by setting all the gauge potentials to be homogeneous in space.
So, the configuration space consists of three copies of 
a Lie algebra. The gauge group acts 
in each copy of the Lie algebra simultaneously in the adjoint
representation.
We take the SO(3)--model discussed in Sec.\ 4 and add two more 
particles ${\bf x}_2$ and ${\bf x}_3$ to the first one ${\bf
x}_1$ ($\equiv {\bf x }$ from Sec.\ 4). Simultaneous rotations of the
position vectors are 
generated by the constraint being 
the sum of all three angular momenta ${\bf l}={\bf
l}_1+{\bf l}_2+{\bf l}_3\approx 0$. These are  three
{\em irreducible}\/ constraints. The \BM treatment of this model is equivalent
to the Faddeev--Popov approach. Indeed, in the gauge $y_1=z_1=z_2=0$ one
obtains  the inner product as the integral  over
the remaining six variables   
with the measure $x_1^2y_2$, which is obviously not
positive definite. The physical wave
functions depend on the six Casimirs ${\bf x}_i\cdot{\bf x}_j,\;1\le
i,j\le 3$ of the model in the gauge chosen. The model suffers from the
Gribov obstruction because the gauge cannot be fixed completely 
\cite{ijmp91}: Here
we have four copies obtained by applying the discrete gauge
transformations $(x_1,x_2)\rightarrow (-x_1,-x_2)$ and $y_2\rightarrow
-y_2$. The physical amplitudes vanish.

So, we conclude that in the presence of a Gribov obstruction (and in
cases where the FP method fails) the reducibility of the constraints 
is crucial for the existence of the global extension of
the \BM procedure. The point which is left and yet to be discussed
is the effect of an infinite number of Gribov copies 
in the \BM inner product. We now  turn to this issue.

\section{The helix model}

In this section we study a model in which the gauge orbits are
(noncompact) helices \cite{helix1,helix2,kl97,fujikawa}. 
The configuration space of the model is a three--di\-men\-sio\-nal 
Euclidean space in which gauge transformations are generated
by the constraint
$G= p_3+x^1p_2-x^2p_1=0$. So they are simultaneous translations 
along the third axis and  SO(2)--rotations in the plane spanned
by $x^{1},x^{2}$. 
The purpose of studying this model is to see what happens to
the \BM inner product if the number of Gribov copies becomes
infinite or may even depend on the position on the gauge
fixing surface. 

Before we proceed, let us make a remark concerning the Gribov
problem in the model. The topology of the gauge orbits in the 
model is that of the real line and thus trivial. There is {\em no}\/
topological obstruction to find a unique single--valued gauge
fixing condition. In fact, e.g.\  the plane $x^3 = 0$ intersects each
gauge orbit precisely once. No Gribov ambiguity occurs in 
contrast to the models with topologically nontrivial gauge
orbits studied above. So the Gribov problem here can be {\em artificially}\/
created by a bad choice of the gauge.\footnote{Such gauges
are easy to find even in electrodynamics. --- ``Artificial
reducibility'' is to be contrasted with what we called essential
reducibility in the Introduction.} For example, with the choice
$x^2=0$ we have infinitely many Gribov copies. Indeed, the plane
$x^2=0$ intersects each helix winding around the third axis
at the points related to one another by transformations
$x^1\rightarrow (-1)^n x^1,\ x^3 \rightarrow x^3 + \pi n$ with
$n$ being any integer. The modular domain on the gauge fixing
surface in configuration space is therefore a half--strip $x^1\geq 0,\
x^3\in [-\pi,\pi)$. Note also that the absence of a global
topological obstruction allows one to construct a well--defined
BRST scalar product in the helix model via the Fock space
representation \cite{fujikawa}. 

The physical amplitudes should not depend on the choice of the gauge.
On the other hand, the \BM inner product explicitly depends on the
BRST gauge fixing fermion. Thus, the independence of physical
amplitudes from the gauge fixing fermion may turn out to be
nontrivial to prove in the presence of a Gribov problem.
We will see that the \BM inner product does not provide 
an interpolation between the two choices of the gauge with
no and an infinite number of Gribov copies, respectively, thus leading to
a general gauge dependence of the physical amplitudes.

The BRST--charge of the model is
\begin{equation}Q=G{\cal C}+\pi\bar{\cal P}\; . 
\end{equation}
The model is irreducible and so it is clear that we have two physical
degrees of freedom. We first take the good gauge $x^3=0$.
The corresponding sets of BRST--doublets read
\begin{equation}D_{(1)}=\{(x^3,{\cal C}),(\bar{\cal C},\pi)\},\;\;  
D_{(2)}=\{({\cal P},G),(\lambda,\bar{\cal P})\}\;.\end{equation}
{}From this 
the hermitian gauge fixing fermion is obtained
\begin{equation}\Psi=x^3\bar{\cal C},\;\;[Q,\Psi]={\cal C}\bar{\cal C}
+ \pi x^3\;.\end{equation}
The auxiliary states are functions of the two Casimirs 
\begin{eqnarray}
C_1 &=& x^1\cos x^3+x^2\sin x^3\ ,\label{c1}\\
C_2&=&x^1\sin x^3-x^2\cos x^3\ .\label{c2}
\end{eqnarray}
The \BM scalar product is now easy to derive
\begin{equation}\<s|s'\>\propto
\int\,dx^1\,dx^2\,\psi^{\ast}(x^1,x^2)\psi'(x^1,x^2)\;.
\end{equation}
Here the arguments of the wave functions are the Casimirs
in the gauge chosen.
The scalar product is well--defined as has been expected 
since the model does not exhibit any topological obstruction
and with $x^3=0$ a good choice of gauge was used.

Let us now calculate the inner product with the bad choice
of gauge $x^2=0$. The sets of BRST--doublets are
\begin{equation}
D_{(1)}=\{(x^2,x^1{\cal C}),(\bar{\cal C},\pi)\},\;\;  
D_{(2)}=\{({\cal P},G),(\lambda,\bar{\cal P})\}\;,
\end{equation}
which lead to the gauge fixing fermion 
\begin{equation}\Psi=x^2\bar{\cal C},\;\;[Q,\Psi]=x^1{\cal C}\bar{\cal C}
+ \pi x^2\;.
\end{equation}
The auxiliary states are given by gauge invariant functions
as above. After simple algebraic computations we find
the \BM inner product
\begin{equation}
\<s|s'\>\propto
\int\,dx^1\,dx^3\,x^1\psi^{\ast}(x^1\cos x^3,x^1\sin x^3)
\psi'(x^1\cos x^3,x^1\sin x^3)\;. 
\end{equation}
It is readily seen that due to the periodicity in $x^3$ of the integrand,
the integral diverges. The periodicity is nothing but the 
residual gauge symmetry in the gauge chosen and the infinite 
factor occurring in physical amplitudes is simply related to
the infinite number of Gribov copies.

One can take an interpolating gauge $\xi x^2 + x^3 = 0$.
When the parameter $\xi$ vanishes we have a good gauge
condition without the Gribov problem and the \BM inner
product is well--defined. The bad gauge is attained when
$\xi$ approaches infinity. Let $\omega$ be a parameter of
the gauge transformations. To obtain the residual gauge
transformations in the gauge chosen, one has to find
all nontrivial values of the 
parameter $\omega$ for which the gauge transformed
configurations belong to the gauge fixing surface.
So we have to solve the system
\begin{eqnarray}
&{}&\xi x^2 + x^3 = 0\ ,\label{xi1}\\
&{}&\xi [x^2 \cos\omega - x^1\sin \omega] + x^3 -\omega = 0\nonumber\ .
\label{xi2}
\end{eqnarray}
For $\xi\neq \infty$, the number of solutions of
this system is finite and so is the number of Gribov copies.
The sets of BRST--doublets in this gauge read 
\begin{equation}
D_{(1)}=\{(\xi x^2+x^3),(\xi x^1+1){\cal C}),(\bar{\cal
C},\pi)\},\;\;   
D_{(2)}=\{({\cal P},G),(\lambda,\bar{\cal P})\}\;.
\end{equation}
For the gauge fixing fermion we find
\begin{equation}
\Psi=(\xi x^2+x^3)\bar{\cal C},\;\;[Q,\Psi]=(\xi x^1+1){\cal
C}\bar{\cal C} 
+ \pi (ax^2+x^3)\;.
\end{equation}
Then the  \BM scalar product becomes 
\begin{equation}
\<s|s'\>\propto
\int\,dx^1\,dx^2\,(\xi x^1+1)\psi^{\ast}\psi'\;, 
\label{ssxi}
\end{equation}
where $\psi,\psi'$ are functions of the Casimirs $C_1,C_2$
in the gauge $x^3 = -\xi x^2$:
\begin{equation}
C_1 = x^1\cos(\xi x^2) + x^2\sin(\xi x^2)\ ,\ \ \ 
C_2 = x^2 \cos(\xi x^2) - x^1\sin(\xi x^2)\ .
\label{c12}
\end{equation}
The integration over the entire plane in (\ref{ssxi}) poses a
problem. The physical states are labeled by the values
of the Casimir functions.
Since the range of values for the variables $x^1,x^2$ is the whole
plane, there is 
no one--to--one correspondence between $(x^1,x^2)$ and $(C_1,C_2)$,
as one may see from (\ref{c12}). For each pair $(C_1,C_2)$,
one can find several pairs $(x^1,x^2)$ that satisfy (\ref{c12}).
All these points on the plane are  Gribov copies of one
another. While the wave functions are invariant under the
Gribov residual gauge transformations, the measure in (\ref{ssxi}) is not.
Moreover the  
number of copies appears to be a function of the point on the gauge
fixing surface.
Therefore for all $\xi\neq 0$, we cannot simply factor out
this number  as in the reducible case, thus shrinking
the integration domain
in (\ref{ssxi}) to the modular domain
(i.e., to the set of $x^1,x^2$ for which (\ref{c12}) is a
one--to--one map).

So we conclude that an infinite number of Gribov copies leads, in
general, to a divergence of the \BM inner product. In the
irreducible case, moreover, the physical amplitude may be gauge dependent,
if the Gribov problem exists in the gauge chosen to construct
the measure of the inner product.   

\section{Conclusion and Outlook}

We have seen that in the case of irreducible constraints
that generate topologically nontrivial gauge orbits
a naive global extension of the \BM inner product encounters
substantial problems and, hence, requires a modification.

In the gauge models with reducible constraints the \BM
construction may apply globally, provided the dependences are
parameterized appropriately within the BRST operator (cf.\ the
discussion in Secs.\ 4 and 5).
It is expected that the positivity
of the measure will not be sufficient for the global
extension in the case of gauge field theories. The reason
is that the number of copies is typically infinite
in the physically interesting gauges. This infinite
number appears as a factor in the \BM inner product.
Though we have made this conclusion from the study
of the helix model where, in fact, no topological
obstruction to the unique gauge fixing exists,
we expect it to be valid for the models where
such an obstruction does exist.

An example for an infinite number of Gribov copies is provided already
by a Yang--Mills theory on a two--dimensional  cylindrical
spacetime (space is compactified into a circle of length $L$)
\cite{migdal,sh93}. The model has a finite number
of physical degrees of freedom which equals the rank
of the gauge group. They can be described by
constant connections taking their values in
the Cartan subalgebra. The residual gauge
transformations that specify the gauge equivalent
configurations in the Cartan subalgebra (Gribov copies)
form the affine Weyl group \cite{sh93}. The latter is a semi-direct
product of the Weyl group $W$ encountered already in Sec.\ 5 and the group
of translations 
in the group unit lattice which consists of Cartan subalgebra
elements whose exponential is the group unit.
This additional gauge translational symmetry makes the
modular domain compact. The modular domain lies in the Weyl chamber
and is called the Weyl cell. For $SU(3)$, the Weyl
cell is an equilateral triangle. As the simplest
example we consider $SU(2)$. The affine Weyl group
consists of reflections on the real line, $x\rightarrow -x$, and
translations, $x\rightarrow x+2nL$, where 
$n$ is an integer. So the only independent
Casimir function is $C= \cos(\pi x/L)$. It is the character
of a group element defined by
the Polyakov loop in the fundamental representation
in the gauge chosen. The number of Gribov copies is infinite.
We have seen that the \BM procedure does not provide
us with a mechanism for the reduction of the integration
domain to the modular domain, only the positivity of the 
measure can be expected. So the infinite number of copies
should appear as a divergent factor in the physical amplitudes. 

In the case of irreducible constraints there seem to
be essentially two obstructions to the global extension
of the \BM inner product. The first one is again a possible
infinite number of copies and the second one is the noninvariance
of the measure under  residual (Gribov) transformations.
The latter problem may lead to a gauge dependence of the 
physical amplitudes or even to their identical vanishing.

Though these conclusions may sound
discouraging because in realistic models none
of these obstructions seem easy to circumvent, we would like to
stress that the possibility of unconventional
gauge fixing fermions has not been explored in our
work. In this respect we would like to mention
recent works \cite{testa} where a modified 
BRST path integral for 
continuous and lattice gauge theories has been proposed to resolve 
the Gribov problem. It might be possible 
to make a similar modification of the \BM procedure
to achieve its global extension.

The problem of the infinite number of copies
in Yang--Mills theory can, in principle, be
circumvented by imposing a gauge condition 
in the momentum space \cite{fsi}.
Since the momenta transform in the adjoint representation,
we can use the gauge fixing procedure from Sec.\ 5 so that
the number of copies would be finite. A rigorous study of
this approach would require a lattice regularization of the
theory in order to give a meaning to spatially local  Weyl
transformations, and this goes beyond the scope of this paper.

Finally, it is worth mentioning that in our paper we have addressed
only kinematical aspects of the global extension of the \BM inner
product. Global obstructions in the BRST formalism occur not only at
the kinematical level (constructing physical states and the proper
inner product), but also on the dynamical level \cite{frsh}. This in
turn may lead to additional restrictions (or conditions) on the
existence of a global extension of the \BM inner product formalism.

\section*{Acknowledgment} 
We are grateful to H.\ Balasin for discussions concerning formula
\re{test}). N.D.\ and T.S.\ want to thank H.\ Kastrup for
encouragement and his interest in our work.  The work of N.D.\ was
partially supported by the DFG under the contract KA-246/16-1. S.V.S.\ 
is grateful to H.\ Kleinert for hospitality at FU--Berlin. He
acknowledges stimulating discussions with F.G. Scholtz and K. Fujikawa
on the early stage of this work and also thanks K. Fujikawa for
sending the preprint \cite{fujikawa}.  The work of S.V.S.\ was
partially supported by the DFG under the contract Kl-25626-1 and the
Alexander von Humboldt foundation.

\end{document}